\documentclass[useAMS,usenatbib]{mn2e}
\usepackage{amssymb,latexsym,amscd,amsmath,graphicx,ifpdf,url}
\usepackage{longtable}

\def\be{\begin{equation}}
\def\ee{\end{equation}}
\def\ba{\begin{array}}
\def\ea{\end{array}}
\def\bi{\begin{itemize}}
\def\ei{\end{itemize}}

\def\half{{\textstyle{1\over2}}}

\newcommand{\bd}{\begin{displaymath}}
\newcommand{\ed}{\end{displaymath}}
\newcommand{\bear}{\begin{eqnarray}}
\newcommand{\eear}{\end{eqnarray}}

\newcommand{\npa}{{\it Nuclear Physics A}}

\newcommand{\prc}{{\it Physical Review C}}

\newcommand{\prl}{{\it Physical Review Letters }}

\newcommand{\apj}{{\it The Astrophysical Journal }}

\newcommand{\apjl}{{\it The Astrophysical Journal Letters }}

\newcommand{\mnras}{{\it Monthly Notices of the Royal Astronomical Society }}

\title[Crust-core coupling during glitches]{Observational constraints on neutron star crust-core coupling during glitches}
\author[W.G.Newton, S. Berger and B. Haskell]{W.G.Newton$^1$, S. Berger$^1$ and B. Haskell$^2$\\
$^1$Department of Physics and Astronomy, Texas A\& M University - Commerce, Commerce, Texas, 75429-3011\\
$^2$School of Physics, The University of Melbourne, Parkville, VIC 3010, Australia}
\begin{document}

\date{\today}

\pagerange{\pageref{000}--\pageref{000}} \pubyear{0000}

\maketitle

\label{firstpage}

\begin{abstract}

We demonstrate that observations of glitches in the Vela pulsar can be used to investigate the strength of the crust-core coupling in a neutron star, and suggest that recovery from the glitch is dominated by torque exerted by the re-coupling of superfluid components of the core that were decoupled from the crust during the glitch. Assuming that the recoupling is mediated by mutual friction between the superfluid neutrons and the charged components of the core, we use the observed magnitudes and timescales of the shortest timescale components of the recoveries from two recent glitches in the Vela pulsar to infer the fraction of the core that is coupled to the crust during the glitch, and hence spun up by the glitch event. Then, within the framework of a two-fluid hydrodynamic model of glitches, we analyze whether crustal neutrons alone are sufficient to drive the glitch activity observed in the Vela pulsar. The analysis is conducted using two sets of neutron star equations of state (EOSs), both of which span crust and core consistently and cover a conservative range of the slope of the symmetry energy at saturation density $30 < L <120$ MeV, a range which encompasses all current experimental constraints. One set produces maximum masses $\approx$2.0$M_{\odot}$, the second $\approx$2.6$M_{\odot}$, allowing us to probe the effect of the high-density stiffness of the EoS as well. We also include the effects of entrainment of crustal neutrons by the superfluid lattice. We find that for medium to stiff EOSs, observations imply $>70\%$ of the moment of inertia of the core is coupled to the crust during the glitch, though for softer EOSs $L\approx 30$MeV as little as $5\%$ could be coupled. No EOS is able to reproduce the observed glitch activity with crust neutrons alone, but extending the region where superfluid vortices are strongly pinned into the core by densities as little as 0.016fm$^{-3}$ above the crust-core transition density restores agreement with the observed glitch activity.

\end{abstract}


\section{Introduction}

Glitches - sudden increases in the rotational frequency of pulsars - offer an insight into the internal dynamics of neutron stars and potential constraints on the properties of dense matter. The quasi-periodic giant glitches exhibited by the Vela pulsar \citep{Melatos:2007gn} offer some of the most stringent tests of our pulsar glitch models, and our models of dense matter \citep{Link:1999ca}. A leading class of glitch model supposes a component of the stellar interior spends most of the time decoupled from the solid crust and magnetosphere whose secular spin-down we track observationally (see \citet{Haskell:2015rev} for a review of glitch models). This component, usually taken to be part of the neutron superfluid, occasionally recouples to the crust and transfers some of its angular momentum, spinning the crust up \citep{Anderson:1975zze,Alpar:1977aa}. The superfluid component can be decoupled by the pinning of vortices to nuclei in the inner crust \citep{Alpar:1977aa,Pines:1980a,Anderson:1982a,Alpar:1984b}, or by pinning to magnetic flux-tubes in the core \citep{Ruderman:1998a,Link:2003hq}. As a rotational lag builds between the decoupled component and the rest of the star, so the hydrodynamic Magnus force builds until it is able to overcome the pinning force and cause the vortices to move outwards collectively, transferring angular momentum to the crust. Due to the complexity of physical glitch scenario, only recently have significant steps been made in a consistent hydrodynamical simulation of the glitch process \citep{Andersson:2005rs,Sidery:2009at,vanEysden:2010ha,Haskell:2011xe,Haskell:2014gl}.

In the most widely examined variant of the model, it is the crustal superfluid that is pinned. However, recent calculations which show that a large fraction of the superfluid neutrons are entrained by the crust via Bragg scattering off the crustal lattice, has led to a vigorous debate on the efficacy of this mechanism in the light of the observed Vela glitch activity \citep{Chamel:2012ae,Chamel:2012zn,Andersson:2012iu,Piekarewicz2014a,Steiner2015a,Newton:2015glitch}. From this, one can conclude that if the whole star spins up during the glitch, it appears that the existence of entrainment in the crust renders the crust-driven glitch scenario only marginally viable at best. If, however, a portion of the core is decoupled from the crust during the glitch, there might be sufficient angular moment stored in the crust superfluid to drive the observed giant glitches \citep{Newton:2015glitch}. 

It is expected quite generally that the core superfluid will be coupled to the normal (non-superfluid) fluid by the mutual friction force. The main mechanism which is thought to be acting in the core is scattering of electrons off superfluid vortex cores, magnetized by entrainment of protons. In this case Mutual Friction will couple the crust and the core superfluid in less than a minute for a typical glitching pulsar with a rotation period of around $10$ ms \citep{Alpar:1984b,Alpar:1988a,Andersson:2005rs}. In the crust, however, protons are not superfluid and electron scattering is ineffective. In this case the Mutual Friction coupling is likely to be much weaker and due mainly to interactions with sound waves in the lattice \citep{Jones1990}. Vortices close to the crust-core boundary, that are mostly immersed in the crust, would thus be weakly coupled, with the coupling becoming stronger at higher densities where vortices are mostly immersed in the core and electron scattering is the dominant Mutual Friction mechanism \citep{Haskell:2011xe}. Comparison with the observational upper limit on the timescale for Vela glitches $<$40s \citep{Dodson:2002gy} suggests that it is likely a significant part of the core will not be coupled to the crust during the glitch. This uncoupled component should then recouple in the minutes after the glitch, a process that should manifest itself observationally in the recovery of the pulsar spin frequency immediately post-glitch \citep{Haskell:2011xe,Haskell:2014gl}. Indeed, in the 2000 and 2004 Vela glitch, exponential fits to the recovery of the spin frequency provide tentative evidence for a short timescale ($\sim$ 1min) component \citep{Dodson:2002gy,Dodson:2006jz}.


Recently, much progress has been made simulating the hydrodynamic evolution of the superfluid vortices including the coupling of crust and core via mutual friction and microscopic pinning forces. Such hydrodynamic models can explain qualitatively the inter-glitch timescales, Vela glitch sizes, and post-glitch rotational evolution, and have the potential to constrain the neutron star equation-of-state (EOS) \citep{vanEysden:2010ha,Haskell:2011xe,Pizzochero:2011dd,Seveso:2012ts}, despite remaining uncertainties in aspects of the glitch model such as the unpinning trigger mechanism and details of how the vortices subsequently unpin \citep{Glampedakis:2008mi,Warszawski:2008fb, Melatos:2009au, Warszawski:2012wa, Warszawski:2012ns, Link2014a}, as well as not including the effect of crustal entrainment.

In this paper, we use the qualitative picture that emerges from hydrodynamic models together with the short timescale component of the 2000 and 2004 Vela glitches to infer the moment of inertia of that portion of the core superfluid that is coupled to the crust during the glitch via mutual friction. We then calculate the maximum moment of inertia of the crustal superfluid neutrons in the crustal regions which drive the glitch according to hydrodynamic models, including the effects of crustal entrainment, and hence infer the glitch activity for the Vela pulsar which we confront with the observed value. The neutron star models used are generated using two families of equations-of-state; one generated using the Skyrme non-relativistic nuclear model and the second using the relativistic mean field (RMF) model. Both families predict the same EOS up to saturation densities, but differ at high densities; the RMF model is stiff and able to accommodate maximum neutron star masses above 2.5$M_{\odot}$, while the Skyrme model used is softer and generally gives maximum masses around 2.0$M_{\odot}$. We generate the families of EOSs by varying the slope of the symmetry energy at saturation density, a quantity that has been shown to correlate strongly with radius and crust-core transition density. We vary it over a conservative range 30-120MeV which encompasses constraints inferred from experiment  \citep{Tsang:2012se,Lattimer:2012xj}. When $L$ is varied, the low-density pure neutron matter is adjusted to maintain a good fit to the results of \emph{ab initio} pure neutron matter (PNM) calculations \citep{Gezerlis:2009iw,Hebeler:2009iv,Gandolfi:2011xu}. In doing so, we are able to explore the predictions of our glitch model over a wide range of EOS parameter space.

In section~2 we explain how we use the observed recovery to set constraints on the fraction of the core coupled to the crust at the time of glitch. In section~3, we apply this to our glitch model using our consistent sets of EOSs. We present our results In section~4, and discuss them in section~5 as well as giving our conclusions.

\section{Crust-core coupling via mutual friction}

The Vela pulsar spins with an angular frequency of $\Omega \approx 70$ rad/s, and exhibits giant glitches $(\Delta \Omega)_{\rm g} \approx 1.5 \times 10^{-4}$ rad/s. The time taken for a glitch to occur (the glitch rise timescale) is constrained by observation to $\tau_{\rm g} \lesssim 40s$. The secular spin-down rate due to electromagnetic torque on the Vela pulsar is $\dot{\Omega} \approx 10^{-10}$ s$^{-2}$ \citep{Dodson:2002gy,Dodson:2006jz}. These are taken to be the values for the solid crust of the Vela pulsar, to which the magnetic field is anchored. The normal component of the star is coupled to the crust on short timescales, but the superfluid neutrons are only weakly coupled, and may not react fast enough to remain coupled during a glitch. Let us take as an illustrative lower limit only 1\% of the total moment of inertia of the star to be coupled to the crust during the glitch. Then, by angular momentum conservation, one would  expect the immediate post-glitch spin-down rate to be $\dot{\Omega} \approx 10^{-8}$ s$^{-2}$.

The key assumption in our model is that the core superfluid recouples post-glitch via mutual friction, in which superfluid vortices become magnetized by entraining protons, allowing electrons to scatter off them and hence couple to the charged components of the core and, through them, the crustal lattice. We assume that the protons and electrons in the core are rigidly rotating and coupled to the crust on short timescales by the magnetic field \citep{Haskell:2011xe, vanEysden:2014}.

Let us now examine how effectively mutual friction couples the crust and core on short timescales. Mutual friction couples the two components on a timescale  \citep{Andersson:2005rs}
\be
\tau_{\rm mf}\sim\frac{1}{2\Omega\mathcal{B}}
\ee
where $\mathcal{B}$ is the mutual friction coefficient \citep{Andersson:2005rs} which expresses the strength of the mutual friction force. Setting this equal to the glitch rise timescale $\tau_{\rm g} \sim 1$min gives us a value of $\mathcal{B}\sim10^{-4}$. All regions of the core with $\mathcal{B}\gtrsim10^{-4}$ will be coupled to the crust during the glitch, while regions of the core with $\mathcal{B}\lesssim10^{-4}$ will begin recoupling to the crust after the glitch, resulting in a torque on the crust. A region with $\mathcal{B}\approx 10^{-5}$, for example, will recouple to the crust on a timescale of several minutes after the glitch, and the resulting torque will give a contribution to the spin down rate, due to mutual friction, of
\be
\dot{\Omega}_{mf}\approx 2\Omega\mathcal{B}\Delta\Omega\approx 10^{-7} \mbox{s$^{-2}$}
\ee
This is considerably stronger than the electromagnetic contribution, even assuming a small amount of the moment of inertia of the core is coupled during the glitch, and we can can infer that core recoupling via mutual friction will give the dominant contribution to the spin-down rate for up to several hours after the glitch. 

%
%
\begin{figure}
\includegraphics[width=6.5cm]{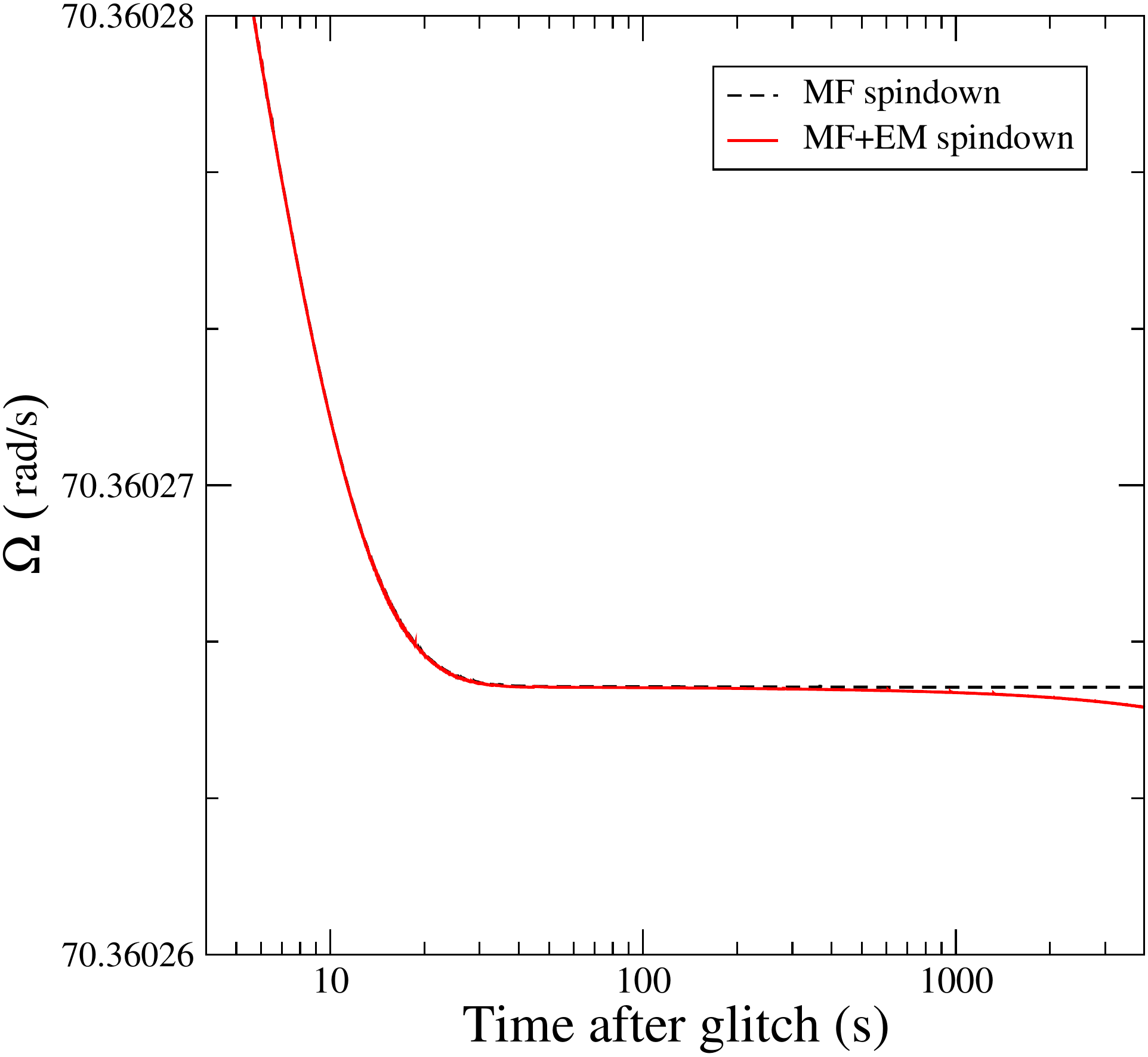}
\caption{The post glitch spin down for the total model (with mutual friction and electromagnetic torques included) and for a model with only mututal friction included. The spin frequency is taken to be that of the Vela pulsar, $\Omega=70$ rad/s. We see that for the first hour or so mutual friction dominates the spin down and the two curves are indistinguishable. After approximately an hour the electromagnetic torque drives the spin down and the curve for the total model begins to show faster spin down.}\label{tscales}
\end{figure}

These estimates are indeed confirmed by running the code of \citet{Haskell:2011xe} for a Vela giant glitch. In Fig (\ref{tscales}) we compare the spin down rate after the glitch obtained for a standard simulation and the rate obtained by excluding the external electromagnetic torque. We can see that the two rates agree in the first few hours in which mutual friction dominates, and only after do they differ as the electromagnetic contribution takes over. Thus immediately after the glitch, the crust component of the star (including the core protons) is spinning faster than a portion of the core neutron superfluid.

If this is the case the observation of short timescale spin recovery after a glitch represents a direct probe of the mutual friction mechanism, and therefore the vortex dynamics in the outer core. We shall now demonstrate how one can infer the moment of inertia of that part of the star coupled to the crust \emph{during} the glitch, $I_{\rm c}$. It is this quantity which enters into the parameter $G = I_{\rm sf}/I_{\rm c}$ (where $I_{\rm sf}$ is the moment of inertia of the superfluid component driving the glitch) to be compared with the pulsar's glitch activity.

\subsection{Extracting the coupled moment of inertia from Vela glitch observations}

There are two published glitches that have an observed short-term recovery component, the Vela 2000 and 2004 glitches \citep{Dodson:2002gy,Dodson:2006jz}. Unfortunately, as shown in table~\ref{vela}, the amplitudes of these components, $\Delta F_{\rm tr}$ vary significantly between the two by more than a factor 1000, partially due to the fact that in the Vela 2004 the short term component was barely above the noise level. The timescales $\tau$ are, however, similar, and consistent with each other within the errors.

%
%
\begin{table}
\begin{tabular}{l l l}
\hline
&2000&2004\\
\hline
$\Delta F_{\rm g}$(Hz) & 2.45435$\times10^{-5}$& 2.2865$\times10^{-5}$\\
$\Delta F_{\rm tr}$ (Hz) & 2$\times10^{-8}$ & 5.4$\times10^{-5}$\\
$\tau$ (min) & 1.2$\pm $0.2 & 1$\pm $0.2\\
\hline
\end{tabular}
\caption{Constant step in frequency $F_{\rm g}$ and fast decaying component of the step $F_{\rm tr}$, with associated timescale $\tau$, for the Vela 2000 and 2004 glitches.} \label{vela}
\end{table} 

Let $I_{\rm n}$ be the moment of inertia of the superfluid region that is in the process of recouping on the timescale $\tau_{\rm mf}$ ($\sim 1$min from table~\ref{vela}), providing the observed short-term contribution to the spin-down of the star, while $I_{\rm c}$ is the moment of inertia of the crust and all components tightly coupled to it during said timescale (which includes the glitch event itself). These are related through \citep{Andersson:2005rs}

\be
\tau_{\rm mf}=\frac{1}{2\Omega\mathcal{B}}\left(\frac{1}{1+I_{\rm n}/I_{\rm c}}\right).
\ee

\noindent We shall refer to these two components as $n$ and $c$.

Before the glitch, the two components will have a small equilibrium lag $(\Delta\Omega)_{\rm eq}=\tau_{\rm mf} \dot{\Omega}\approx 10^{-8}$ for the Vela pulsar.
The glitch acts to spin up the component $c$ by an amount $(\Delta \Omega)_{\rm g} \gg  (\Delta\Omega)_{\rm eq}$, and therefore immediately after the glitch the lag between the components will be  $(\Delta \Omega)_{\rm in} = (\Delta \Omega)_{\rm g}$. After the timescale over which the component $n$ recouples, component $c$ has decreased its frequency by an amount $(\Delta \Omega)_{\rm tr}$ and component $n$ has increased its frequency by $(\Delta \Omega)_{\rm g}  - (\Delta \Omega)_{\rm tr}$; thus by angular momentum conservation

\be \label{tmf}
\frac{I_{\rm n}}{I_{\rm c}}=\frac{ (\Delta\Omega)_{\rm tr}}{ (\Delta\Omega)_{\rm g}-(\Delta\Omega)_{\rm tr}}
\ee

\noindent Taking $(\Delta \Omega)_{\rm g}$ and  $(\Delta\Omega)_{\rm tr}$ from observations, we obtain
\bear
\frac{I_n}{I_c}\approx 8\times 10^{-4}\;\;\;\mbox{for the Vela 2000 glitch}\\
\frac{I_n}{I_c}\approx 2.37 \;\;\;\mbox{for the Vela 2004 glitch}
\eear

Using this result and the observed timescales, equation~\ref{tmf} gives us values for the mutual friction parameter above which correspond to regions of the core coupled at the time of glitch

\bear
\mathcal{B}=1\times 10^{-4} \pm 3\times 10^{-5}\;\;\;\;\mbox{for the Vela 2000 glitch}\\
\mathcal{B}=4\times 10^{-5} \pm 1\times 10^{-5}\;\;\;\;\;\;\;\;\mbox{for the Vela 2004 glitch}
\eear

These numbers are still (roughly) consistent given the uncertainties that are clearly underestimated in this procedure. In the following we shall simply assume a range:
\be
 3\times 10^{-5} < \mathcal{B}_{\rm obs} < 1.3\times 10^{-4}.
 \label{range}
\ee

\noindent from which we can compute the moment of inertia of the core $I_{\rm c}$ coupled via mutual friction to the crust at the time of glitch, and hence spun up by the glitch. This procedure, although limited by current observational uncertainties, allows us to derive this important quantity from observations for the first time.

We now outline how we use this result to extract $I_{\rm c}$ from neutron star models.

\section{Modeling the glitch}

Given a particular neutron star equation of state (EOS), we obtain our background neutron star model by solving the Oppenheimer-Volkoff (OV) equations, thus obtaining the density and composition profile of the star as a function of radial coordinate $r$.

The moment of inertia of a star of radius $R$ in the limit of small angular frequency $\Omega$ \citep{Hartle:1968si} is given by

\begin{equation} \label{eq:MoI1}
I_{\rm tot}=\frac{8 \pi}{3} \int_0^R r^{4}e^{-\nu(r)}\frac{\bar{\omega}(r)}{\Omega}\frac{\left( \mbox{\Large{$ \varepsilon $}} (r)+P(r) \right)}{\sqrt{1-2GM(r)/r}}\mathrm{d}r,
\end{equation}

\noindent where $\text{\Large{$ \varepsilon $}}(r)$ is energy density of matter in the star, $P(r)$ is the pressure and $M(r)$ is the mass contained in radius $r$. $\nu(r)$ is a radially-dependent metric function given by

\begin{equation} \label{eq:MoI2}
\nu(r)=\frac{1}{2} \ln \left( 1- \frac{2GM}{R} \right) - G \int_r^R \frac {\left( M(x)+4 \pi x^{3} P(x) \right)}{x^{2} \left( 1 - 2 GM(x)/x \right)} \mathrm{d}x,
\end{equation}

\noindent and $\bar{\omega}$ is the frame dragging angular velocity

\begin{equation} \label{eq:MoI3}
\frac{1}{r^{3}}\frac{\mathrm{d}}{\mathrm{d}r} \left( r^{4}j(r)\frac{\mathrm{d}\bar{\omega}(r)}{\mathrm{d}r} \right) + 4 \frac{\mathrm{d}j(r)}{\mathrm{d}r} \bar{\omega}(r)=0,
\end{equation}

\noindent where

\begin{equation} \label{eq:fp3}
j(r)=e^{-\nu(r)-\lambda(r)}=\sqrt{1-2GM(r)/r} e^{-\nu(r)}
\end{equation}

\noindent for $r \leq R$. 

We take the standard form for the mutual friction coefficient in the core, due to electron scattering off magnetized vortex cores \citep{Andersson:2005rs}:
\bear \label{eqn:B}
\mathcal{B}&=&4\times 10^{-4} \left(\frac{m_p-m_p^*}{m_p}\right)^2\left(\frac{m_p}{m_p^*}\right)^{1/2}\times \notag \\
&&\times\left(\frac{x_p}{0.05}\right)^{7/6}\left(\frac{\rho}{10^{14}\mbox{g/cm$^3$}}\right)^{1/6}
\eear
\noindent where $x_{\rm p}$ is the proton fraction, $m_{\rm p}^*$ the microscopic effective proton mass and $\rho$ the total density in the core. These quantities are obtained consistently from the EOS, as a function of radial coordinate $r$ once the OV equations are solved.  As detailed in \citet{Haskell:2011xe}, we average $\mathcal{B}$ over the $z$-direction to find the average mutual fiction strength experienced by a vortex at cylindrical radius $\tilde{r}=r\sin\theta$.
\be
\bar{\mathcal{B}}(\tilde{r})=\frac{\int^{\theta(\tilde{r})}_{\pi/2}\frac{2\mathcal{B}}{\sqrt{1-2M(r)/r}}r d\theta}{ \int^{\theta(\tilde{r})}_{\pi/2}\frac{2}{\sqrt{1-2M(r)/r}}r d\theta}
\ee

We can now define a region of the star bounded by cylindrical radii $\tilde{r}_1$ and $\tilde{r}_2$ such that $\bar{\mathcal{B}}(\tilde{r}_{1} < \tilde{r} < \tilde{r}_2) > \mathcal{B}_{\rm obs}$. This corresponds to the region in which the superfluid neutrons are strongly coupled to the crust on the glitch rise time. 

%
%
\begin{figure}\label{fig:1}
\begin{center}
\includegraphics[width=6cm,height=6cm]{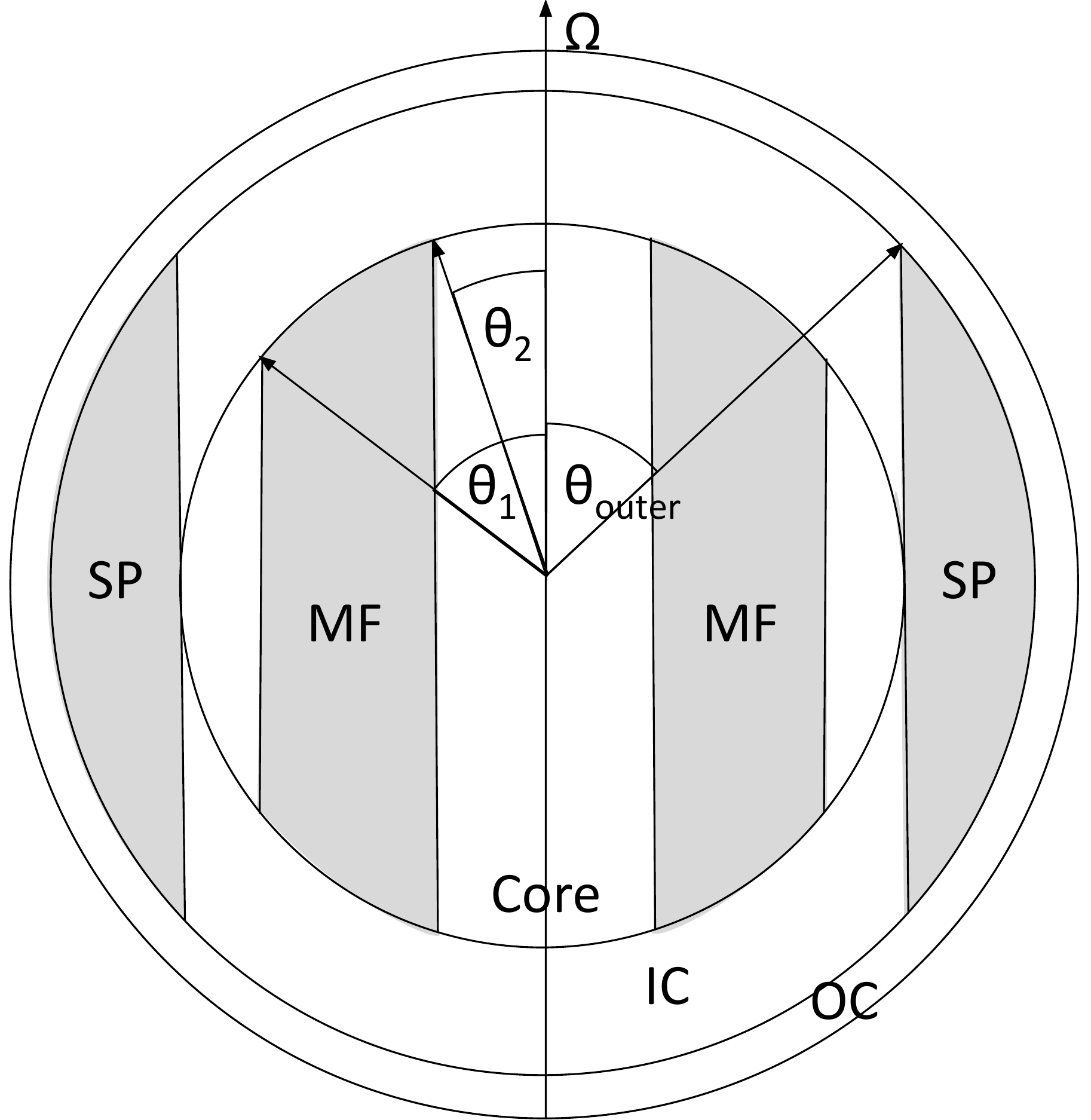}
\caption{Neutron star cross-section in plane of rotation axis ($\Omega$) depicting the geometry of the strong pinning region in inner crust (shaded area SP) and the region of the core coupled to the crust at the time of the glitch via mutual friction (shaded region MF). OC and IC label the outer and inner crust respectively. The angular locations of the boundary of the strong pinning region at the outer crust, and the outer and inner boundaries of the coupled region of the core, are given by $\theta_{\rm outer}, \theta_1$ and $\theta_2$ respectively. Except for the softest equations of state at saturation density (L $\approx$ 30 MeV), $\theta_2$ = 0.}
\end{center}
\end{figure}

Defining

\begin{equation} \label{eq:new2}
r^{2} \mathcal{I} =\frac{8 \pi}{3} r^{4}e^{-\nu(r)}\frac{\bar{\omega}(r)}{\Omega}\frac{\left( \mbox{\Large{$ \varepsilon $}}_{n} (r)+P_{n}(r) \right)}{\sqrt{1-2GM(r)/r}},
\end{equation}

\noindent where $\mbox{\Large{$ \varepsilon $}}_{n} (r)$ is the energy density of the superfluid neutrons, $P_{n}(r)$ is the pressure of the superfluid neutrons,
then the moment of inertia contained in the regions of the star with cylindrical radius greater than $\tilde{r}_1$ and $\tilde{r}_2$ respectively is given by

\be
I_{1,2}=\int^{\pi/2}_{\theta_{1,2}}\int^{R_{\rm cc}}_{R(\theta)} r^2\mathcal{I}\sin\theta d\theta dr
\ee

\noindent and therefore the moment of inertia of the region in which the superfluid neutrons are strongly coupled to the crust on the glitch rise time is given by

\be
I_{\rm nc} = I_2 - I_1.
\ee

%
%
\begin{figure*}\label{fig:MR}
\begin{center}
\includegraphics[width=8cm,height=6cm]{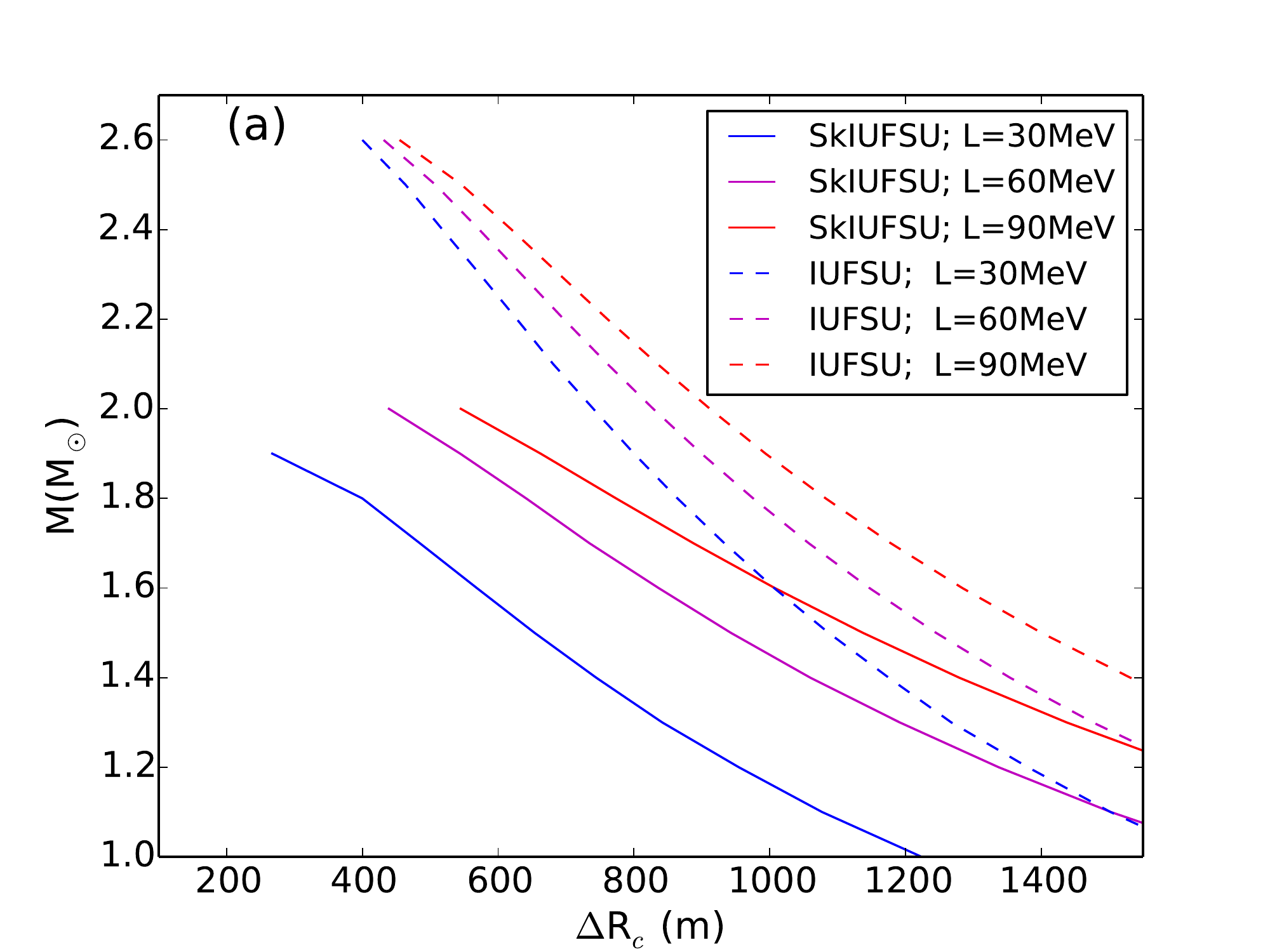}\hspace{0.5cm}\includegraphics[width=8cm,height=6cm]{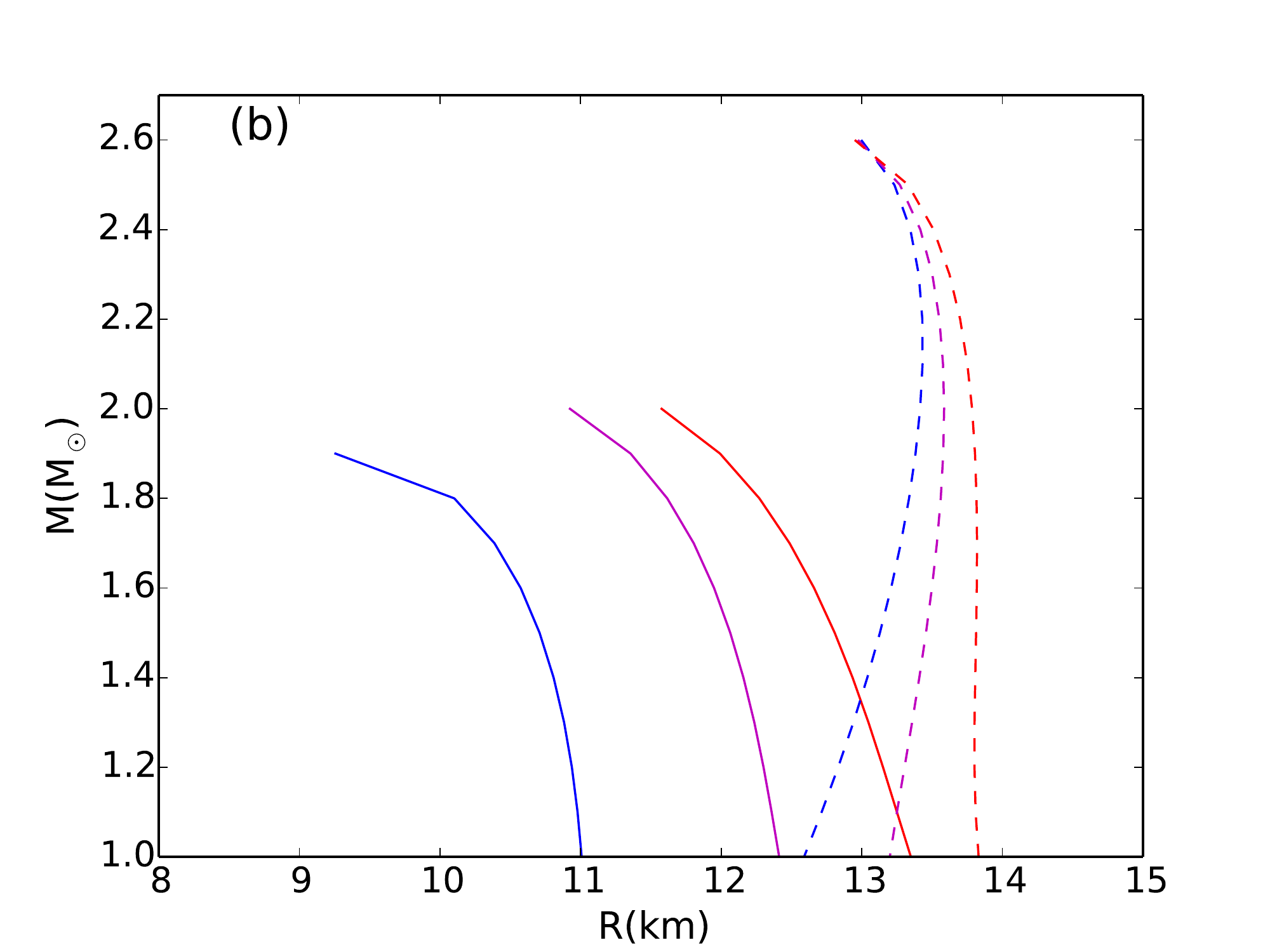}
\caption{(Color online) Crust thickness $\Delta R_{\rm c}$ (a) and  Total radius $R$ (b) versus mass, for the $L=30$ (blue), $60$ (pink) and $90$MeV (blue) members of the non-relativistic SkIUFSU EOS which is softer at high densities (solid lines) and the relativistic IUFSU EOS (dashed lines) which is stiffer at high densities.}
\end{center}
\end{figure*}

\noindent Here, $R_{\rm cc}$ is the radius of the crust-core boundary, and $\theta_1, \theta_2$ are the angular locations where the outer and inner boundaries of the coupled region where it meets the crust-core boundary respectively (see~Fig. 2). The total moment of inertia of that part of the star strongly coupled to the crust at glitch rise time is obtained by adding the contribution of the crustal lattice itself and the protons in the core

\be
I_{\rm c} = I_{\rm nc} + I_{\rm crust} + I_{\rm p}.
\ee

Hydrodynamic simulations of vortex evolution suggest that  only the crustal superfluid neutrons within the strong pinning region of the crust, defined as the region within which vortices are totally immersed in the inner crust, contribute to the glitch itself. How many contribute depends on details of the pinning force throughout the crust \citep{Haskell:2011xe}, but here we take the entirety of the strong pinning region as an upper limit. The moment of inertia of the strong pinning region of inner crust superfluid neutrons is

\be \label{eq:new3}
I_{\rm csf}^{\rm (sp)}=\int_{\theta_{\rm outer}}^{\pi/2} \int^{R(\theta_{\rm outer})}_{R(\theta)} r^{2} \mathcal{I}\mathrm{d}r \sin \theta \mathrm{d} \theta
\ee

\noindent where $R(\theta)$ is the distance from the core of the star to the inner boundary of the strong pinning region at an angle $\theta$ to the rotation axis, $R(\theta_{\rm inner}) \equiv R_{\rm inner}$ and $R(\theta_{\rm outer}) \equiv R_{\rm outer}$ (see Fig.~2). 

Entrainment of superfluid neutrons by the crust's lattice reduces the mobility of the neutrons with respect to that lattice. It can be shown that this effect is encoded by introducing an effective \emph{mesoscopic} neutron mass $m_{\rm n}^*$ \citep{Chamel:2004in,Chamel:2005my,Chamel:2012zn}; larger values correspond to stronger coupling between the neutron superfluid and the crust, and a reduction in the fraction of superfluid neutrons able to store angular momentum for the glitch event. One can include this effect by modifying the integrand Eq.~\ref{eq:new2} in the inner crust:

\be
r^{2} \mathcal{I} \to r^{2} \mathcal{I}^* = {m_{\rm n} \over m_{\rm n}^* (r)} r^{2} \mathcal{I}
\ee

\noindent where $m_{\rm n}^*(r)$ is the effective mass at radius $r$ in the crust. We obtain $m_{\rm n}^*(r)$ from the results of Chamel \citep{Chamel:2012zn} by interpolating between the values calculated at specific densities to find the effective mass at arbitrary locations in the inner crust.

%
%
\begin{figure*}\label{fig:B}
\begin{center}
\includegraphics[width=8cm,height=6cm]{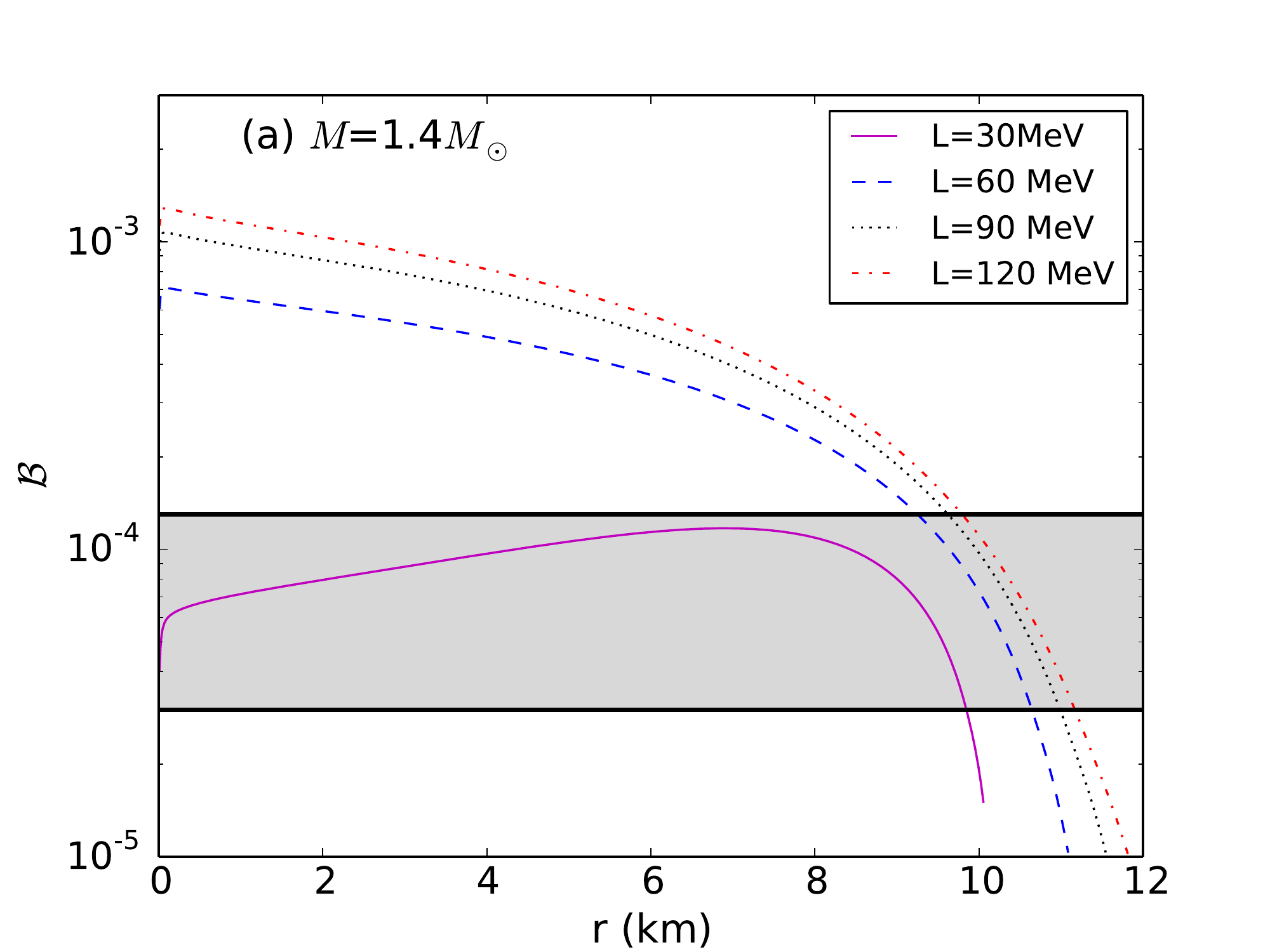}\hspace{0.5cm}\includegraphics[width=8cm,height=6cm]{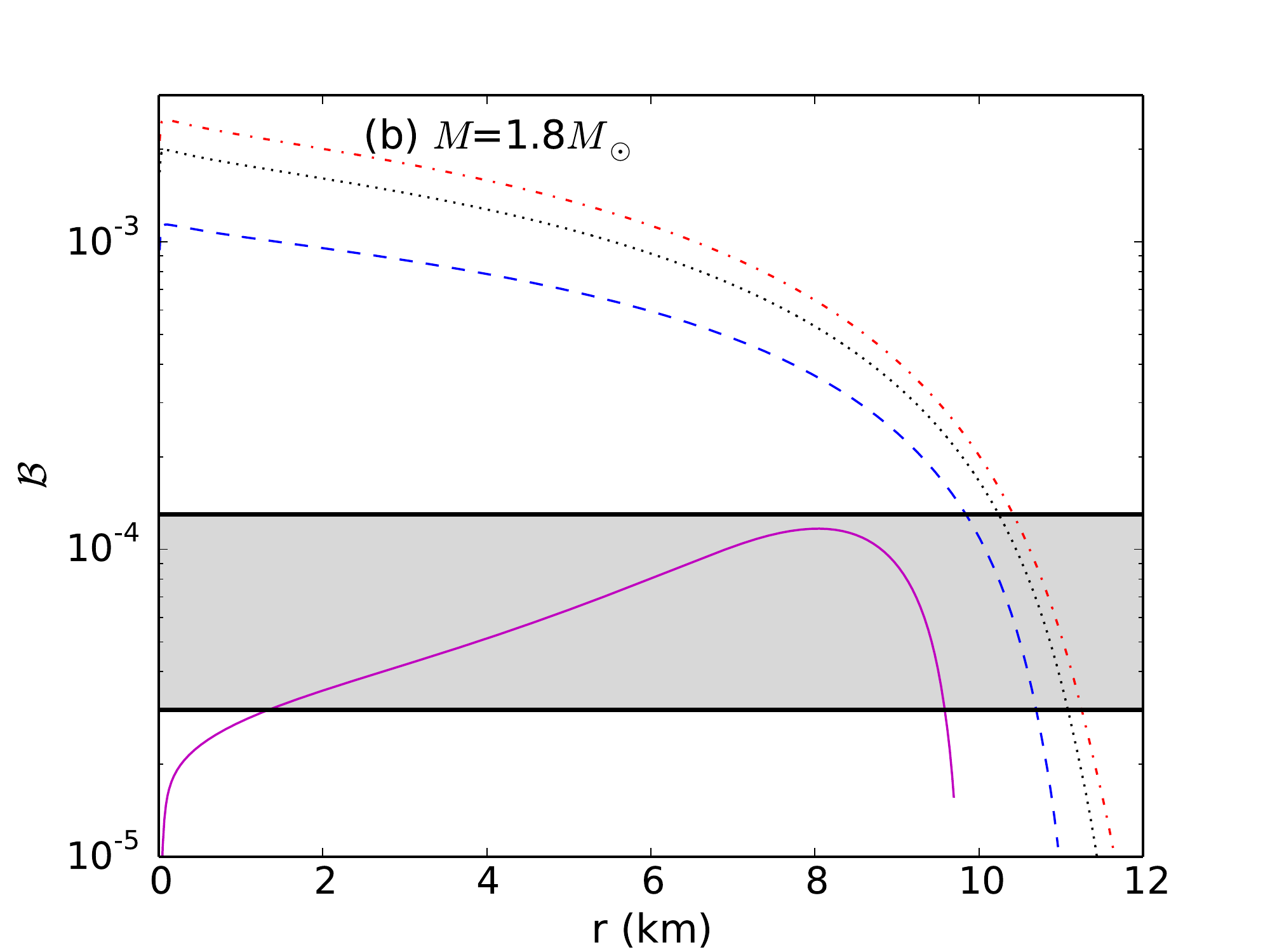}
\caption{(Color online) Mutual friction profiles as a function of radial coordinate for a 1.4 $M_{\odot}$ (a) and 1.8$M_{\odot}$ (b) star for the $L$=30,60,90 and 120 MeV members of the SkIUFSU EOS family. We infer from observations the value above which the superfluid neutrons are tightly coupled to the crust; the shaded region depicts the upper and lower bounds of that value.}
\end{center}
\end{figure*}

\subsection{Nuclear matter parameters and crust and core equations of state}

The glitch model requires several microscopic quantities as input. These include the total pressure and energy density $P(n_{\rm b})$, $\Large{\varepsilon} (n_{\rm b})$, neutron pressures and energy densities  $P_{\rm n} (n_{\rm b})$, $\Large{\varepsilon}_{\rm n} (n_{\rm b})$, effective proton mass in the core,  proton fraction and mass density $m^*_{\rm p}(n_{\rm b})$, $x_{\rm p}(n_{\rm b})$ and $\rho (n_{\rm b})$, as a function of baryon number density $n_{\rm b}$, and the crust-core transition density and pressure $n_{\rm cc}$ and $P_{\rm cc}$. These quantities are derived from an underlying model of nuclear matter.

Experimental information about nuclear matter is predominantly extracted from nuclear systems at densities around nuclear saturation density $n_0 = 0.16$ fm$^{-3}$ and at proton fractions close to one half. As a consequence, nuclear matter models are generally characterized by their behavior in that region of parameter space. Denoting the energy per particle of nuclear matter around saturation density by $E(n_{\rm b},\delta)$ where $\delta = 1-2x_{\rm p}$ is the isospin asymmetry parameter; $\delta = 0$ corresponds to symmetric nuclear matter (SNM), and $\delta = 1$ to pure neutron matter (PNM). The \emph{nuclear symmetry energy} $S(n)$ is defined as the quadratic coefficient in the expansion of $E(n_{\rm b},\delta)$ about $\delta = 0$:

Nuclear matter models can be characterized by their behavior around nuclear saturation density $n_0 = 0.16$ fm$^{-3}$, the density region from which much of our experimental information is extracted. We can denote the energy per particle of nuclear matter around saturation density by $E(n_{\rm b},\delta)$, where $n_{\rm b}$ is the baryon density and $\delta = 1-2x$ the isospin asymmetry, where $x$ is the proton fraction. $x=0.5, \delta = 0$ corresponds to symmetric nuclear matter (SNM), and $x=0, \delta = 1$ to pure neutron matter (PNM). By expanding $E(n, x)$ about $\delta = 0$ we can define the \emph{symmetry energy} $S(n)$,

\be\label{eq:eos1}
E(n,\delta) = E_{\rm 0}(n_{\rm b}) + S(n_{\rm b})\delta^2 + ...,
\ee
\noindent The symmetry energy is the energy cost of increasing the isospin asymmetry of matter and is a function of baryon density. Furthermore, we can expand  the symmetry energy about saturation density using the density parameter $\chi = \frac{n_{\rm b}-n_{\rm 0}}{3n_{\rm 0}}$:
\be\label{eq:eos3}
	S(n_{\rm b}) = J + L \chi + \half K_{\rm sym} \chi^{2} + ..., 
\ee

\noindent where $J$, $L$ and $K_{\rm sym}$ are the symmetry energy, its slope and its curvature at saturation density. Over the past decade, vigorous effort has been devoted to experimentally constraining $J$ and particularly $L$ \citep{Li:2008gp,Tsang:2012se}. Currently the congruence of experimental results \citep{Hebeler:2013nza} points to the range  $30<L<60$ MeV, but stiffer (higher values of $L$) are not conclusively ruled out \citep{Fattoyev:2013lm}; we therefore examine a conservative range of $30<L<120$ MeV in this paper.

Crust and core EOSs and the transition density are calculated consistently using two models of the nuclear many-body interaction. We use the IUFSU parameterization of the relativistic mean-field model \citep{Fattoyev:2010mx} and a parameterization of the non-relativistic Skyrme model, SkIUFSU, which shares the same saturation density symmetric nuclear matter (SNM) properties as IUFSU, used in previous work \citep{Fattoyev:2012ch,Fattoyev:2012uu,Newton:2015glitch}. Both models have isovector nuclear matter parameters obtained from a fit to state-of-the-art PNM calculations  \citep{Gezerlis:2009iw,Hebeler:2009iv,Gandolfi:2011xu}, which makes them particularly suitable for describing the low-density neutron fluid in the inner crust, and both describe the bulk properties of doubly magic nuclei well \citep{Fattoyev:2012ch}. The presence in both models of two purely isovector model parameters allows the density slope of the symmetry energy $L$ to be systematically adjusted, while retaining the fit to PNM at low densities. The fact that they are isovector means that such adjustments leave SNM properties unchanged \citep{Chen:2009wv}. This adjustment under the PNM constraint leads to the relation $J = 0.167 L + 23.33$ MeV. 

Note that you generally cannot separate out the individual proton and neutron pressures in these nuclear matter models - the pressure contains terms that are not separable into neutron and proton components. However, in both Skyrme and RMF models, the enthalpy density $h = \Large{\varepsilon} + P$ is separable, because the non-separable terms in the pressure and energy density cancel out. It is the enthalpy that appears in the moment of inertia integrals.

Both nuclear matter models give closely identical EOSs up to saturation density for a given value of $L$. The SkIUFSU model is softer than IUFSU at high densities and but gives maximum masses of $M \approx 2$M$_{\odot}$ for all values of $L$, matching the observational lower limit \citep{Demorest:2010bx,Antoniadis:2013}. The IUFSU model we have chosen to be maximally stiff at high densities by adjusting the parameter $\zeta$ of the RMF model that controls the quartic
omega-meson self interactions~\citep{Mueller:1996pm} and subsequently the high-density component of the EoS of SNM. It is set so that the SNM EoS is maximally stiff, resulting in maximum mass neutron stars of $\gtrsim 2.5M_{\odot}$. The crustal EoS and crust-core transition densities are derived from the compressible liquid drop model (CLDM) for the crust \citep{Newton:2011dw} using the same nuclear matter model as the core EoS. EoS quantities obtained include the proton fraction $x_{\rm p}$ and microscopic effective proton mass $m_{\rm p}^*$ as a function of density, required to calculate the mutual friction parameter $\mathcal{B}$ from equation~\ref{eqn:B}. For a 1.4 $M_{\odot}$ star, the softest EoS $L=$30 MeV produces effective proton masses that vary from 0.7 down to 0.3 from crust to core and the stiffest EoS $L=$120 MeV produces effective proton masses that vary from 0.85 down to 0.15 from crust to core.

%
%
\begin{figure*}\label{fig:Ic/It}
\begin{center}
\includegraphics[width=8cm,height=6cm]{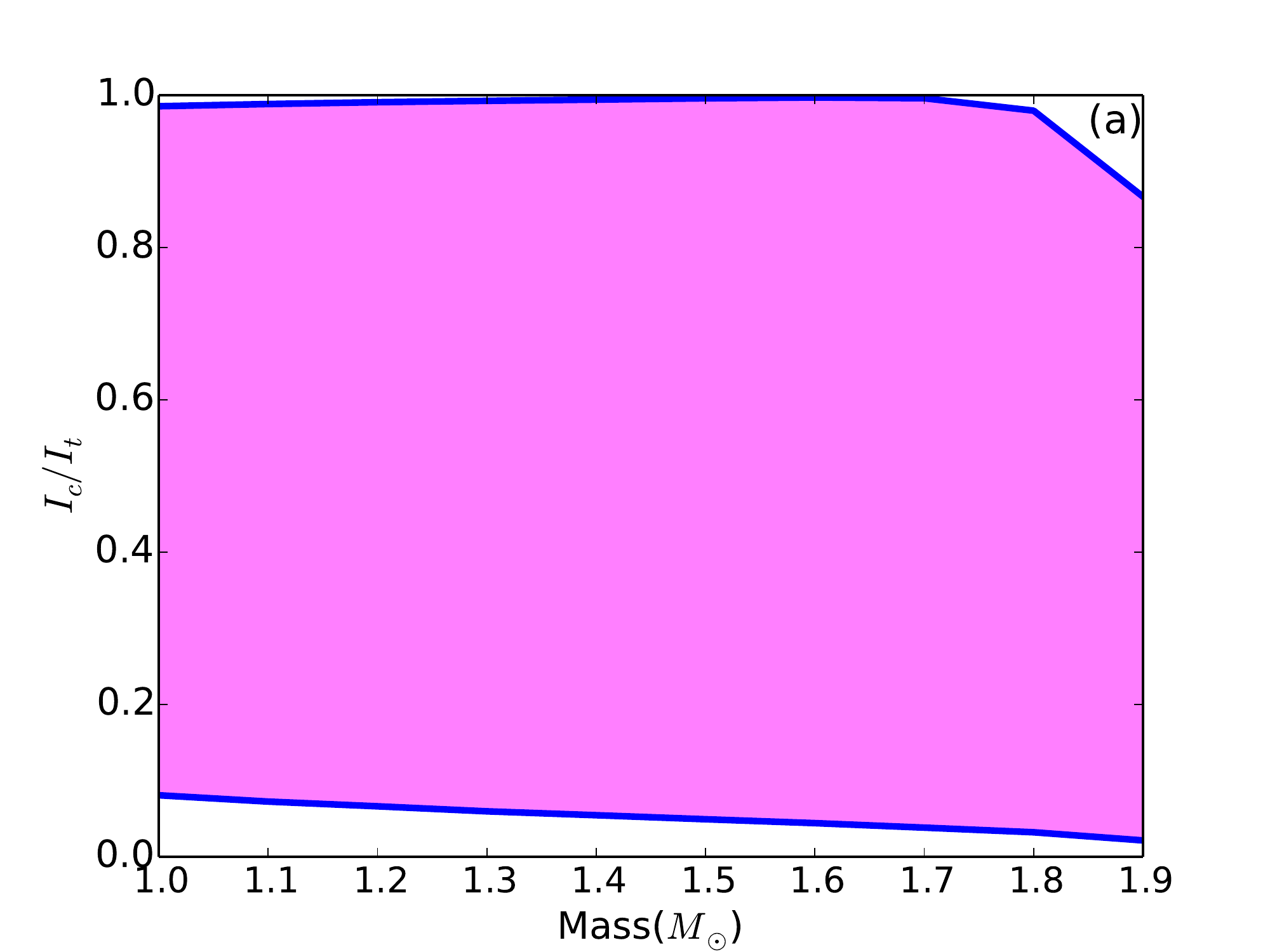}\hspace{0.5cm}\includegraphics[width=8cm,height=6cm]{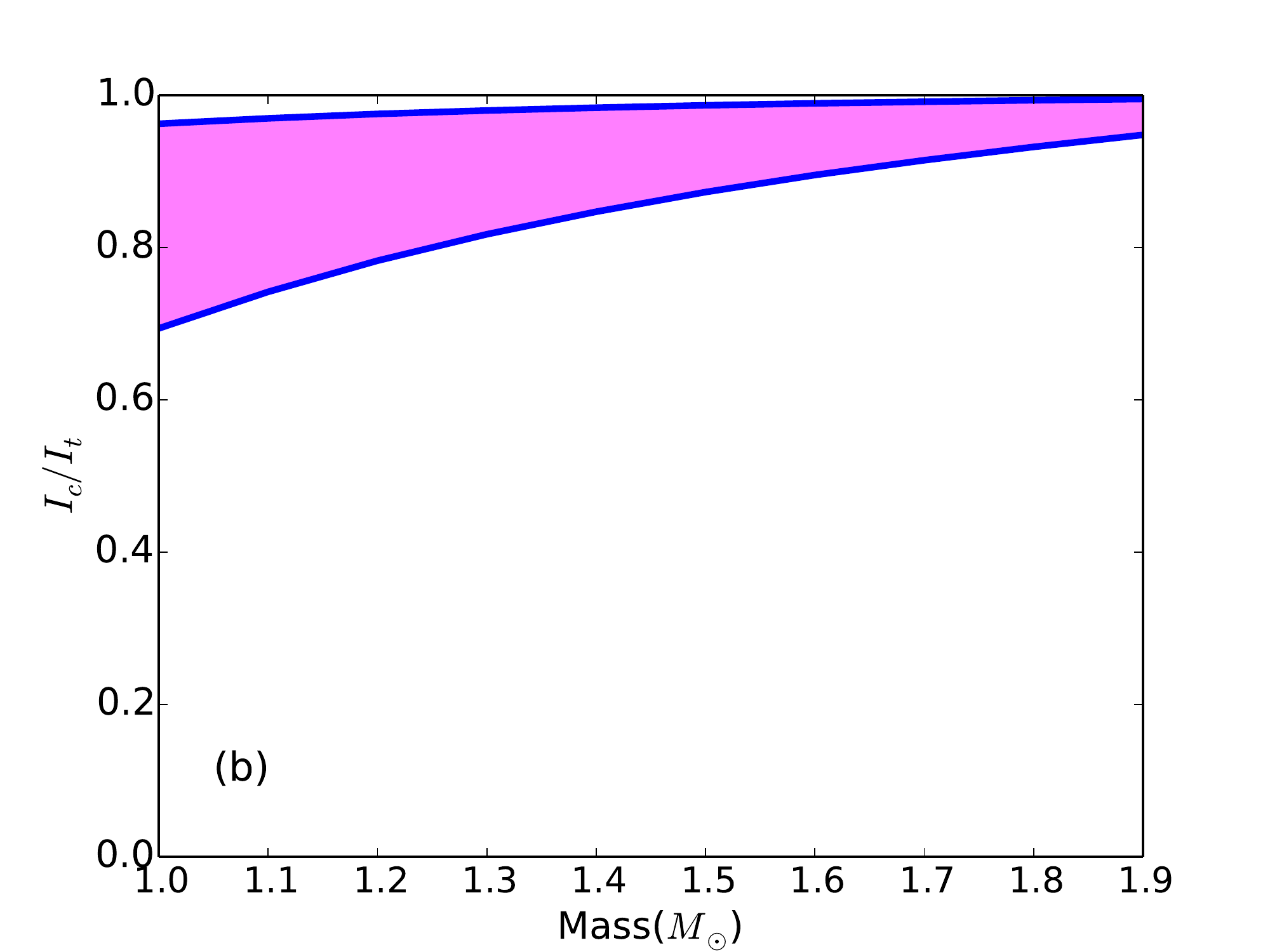}
\includegraphics[width=8cm,height=6cm]{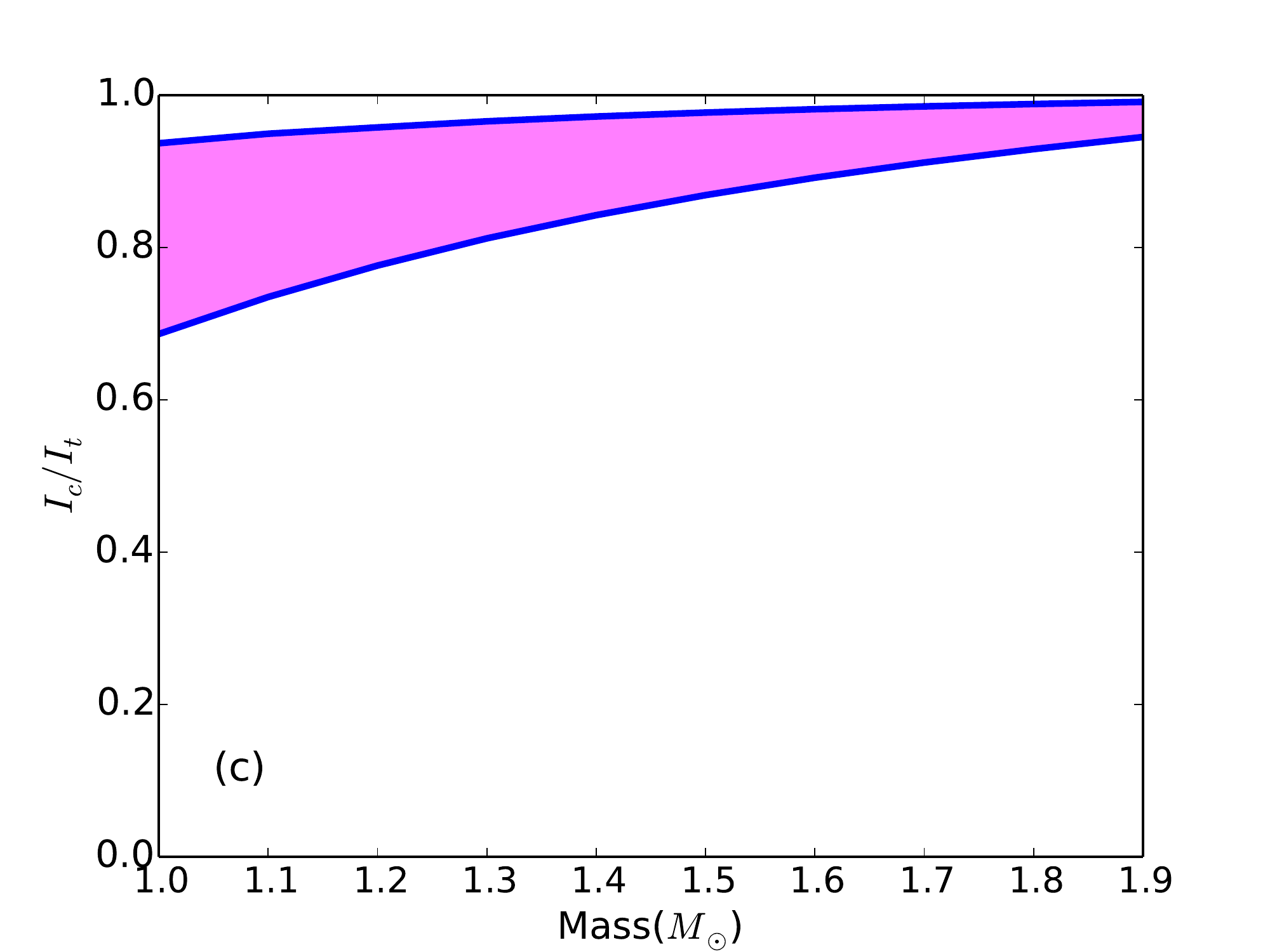}\hspace{0.5cm}\includegraphics[width=8cm,height=6cm]{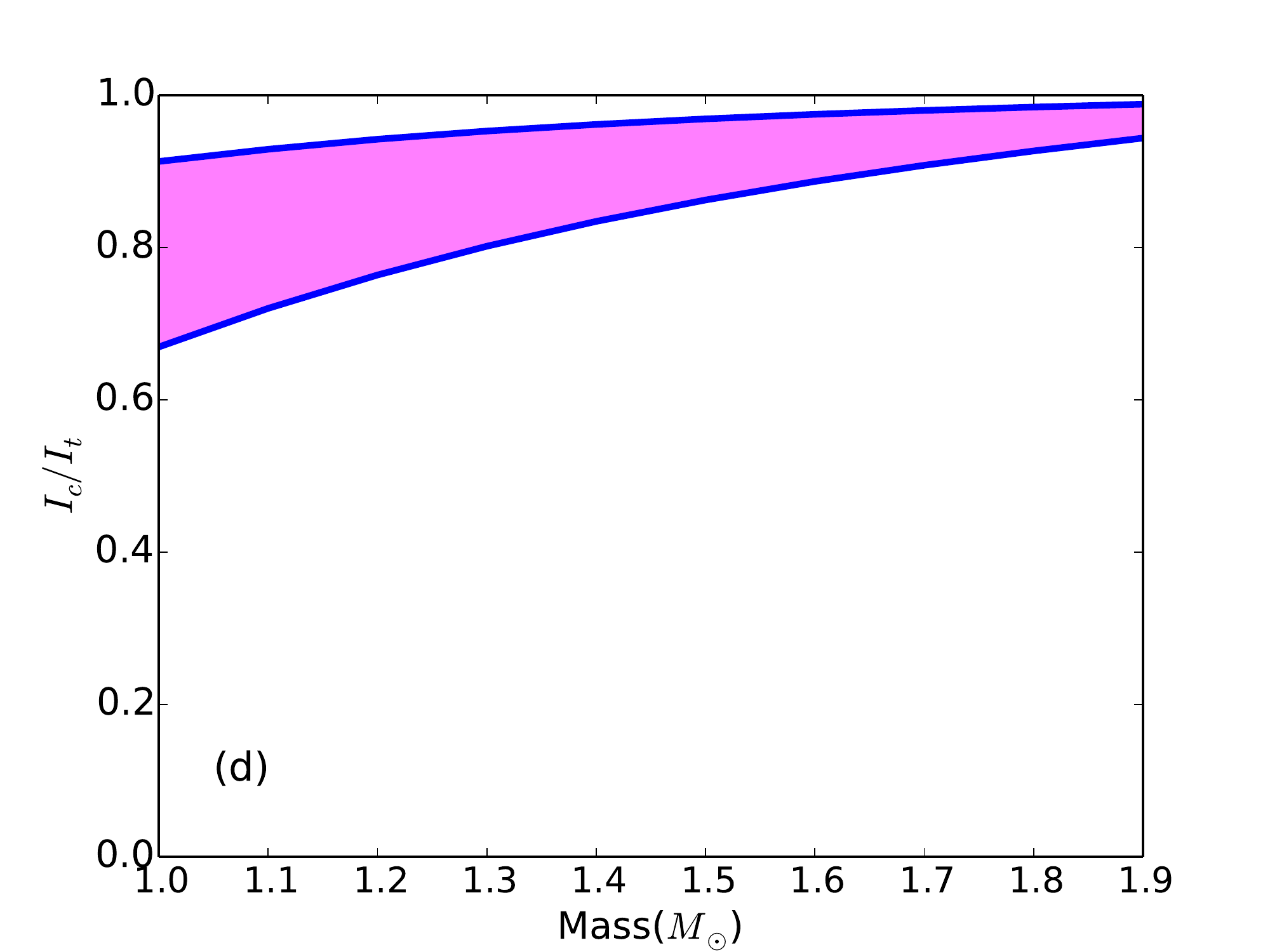}
\caption{Fractional moment of inertia of the component of the star coupled to the crust during the glitch $I_{\rm c}/I_{\rm tot}$ as a function of stellar mass for$L=30$MeV (a), $L=60$MeV (b), $L=90$MeV (c) and $L=120$MeV (d). The ranges span the upper and lower bounds inferred from observation, and correspond to taking the upper and lower bounds of the shaded regions in figure~\ref{fig:B}.}
\end{center}
\end{figure*}

In Fig.~\ref{fig:MR} we show the bulk properties of neutron stars for three representative values of the slope of the symmetry energy $L=30, 60$ and $90$ MeV for the two families of EOSs: SkIUFSU (maximum masses around 2 M$_{\odot}$) and IUFSU (maximum masses around 2.5-2.6 M$_{\odot}$). In the left panel, the thickness of the crust $\Delta R_{\rm c}$ is displayed as a function of mass $M$, while on the right the total radius of the star $R$ is displayed as a function of mass $M$. 

\section{Results}

Before we get into discussions of the results, we remind readers that when we refer to the portions of the core coupled and uncoupled to the crust, we are referring specifically to the superfluid core neutrons. The small proton component of the core is assumed to always be tightly coupled to the crust.

In Fig.~4 we show the averaged mutual friction profile $\bar{\mathcal{B}}(\tilde{r})$ as a function of cylindrical radius $\tilde{r}$ throughout the core of a 1.4 M$_{\odot}$ star (a) and a 1.8 M$_{\odot}$ star (b). In each case we show the mutual friction profiles obtained with the $L$=30, 60, 90 and 120MeV members of the SkIUFSU EOS family. The shaded band covers the possible range of the threshold strength of mutual friction inferred from the 2000 and 2004 Vela glitches. All regions of the star with a mutual friction greater than this threshold are coupled to the crust at the time of glitch, while those regions below the threshold are uncoupled from the crust at the time of glitch.

%
%
\begin{figure*}\label{fig:Is}
\begin{center}
\includegraphics[width=8cm,height=6cm]{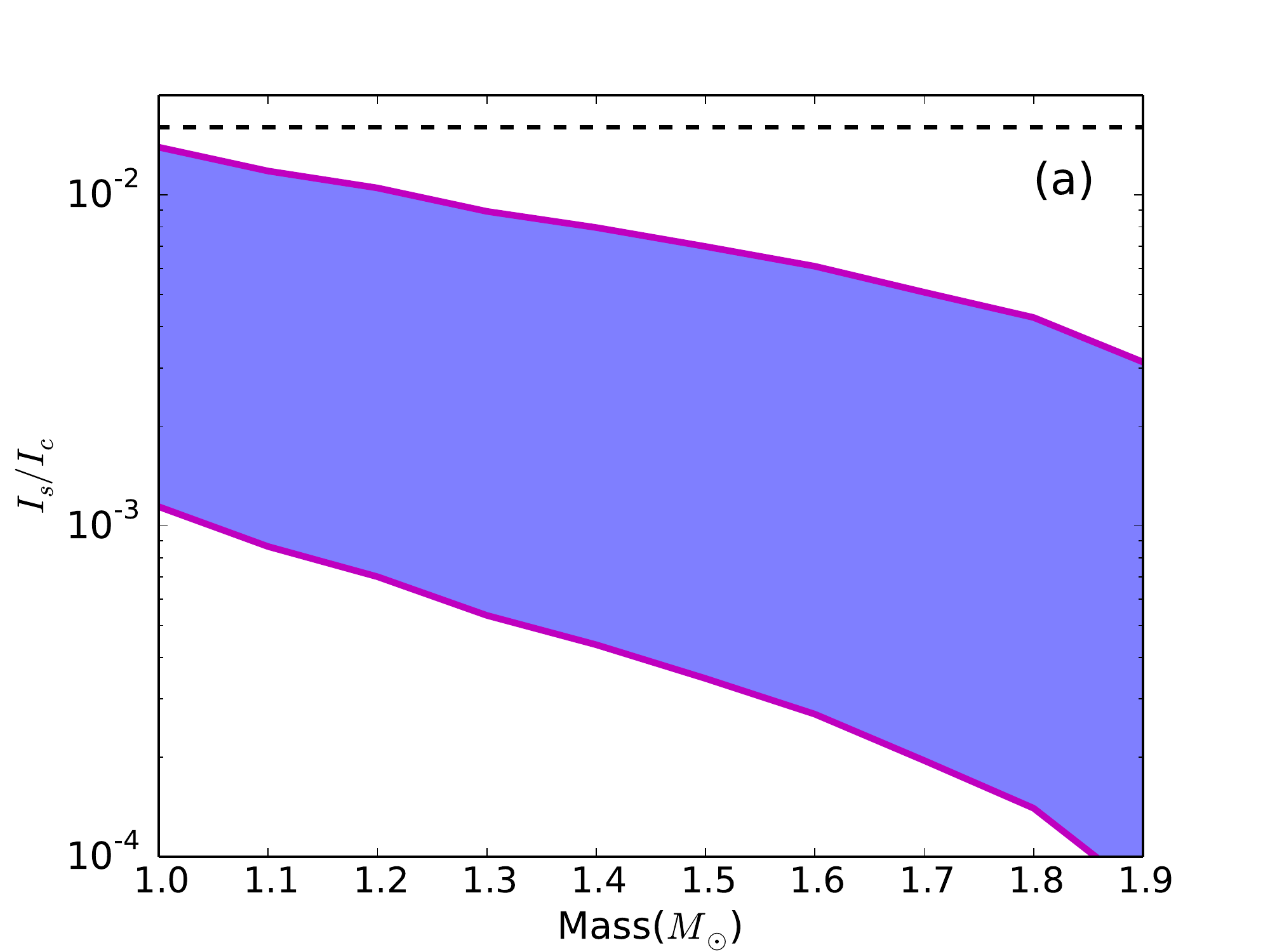}\hspace{0.5cm}\includegraphics[width=8cm,height=6cm]{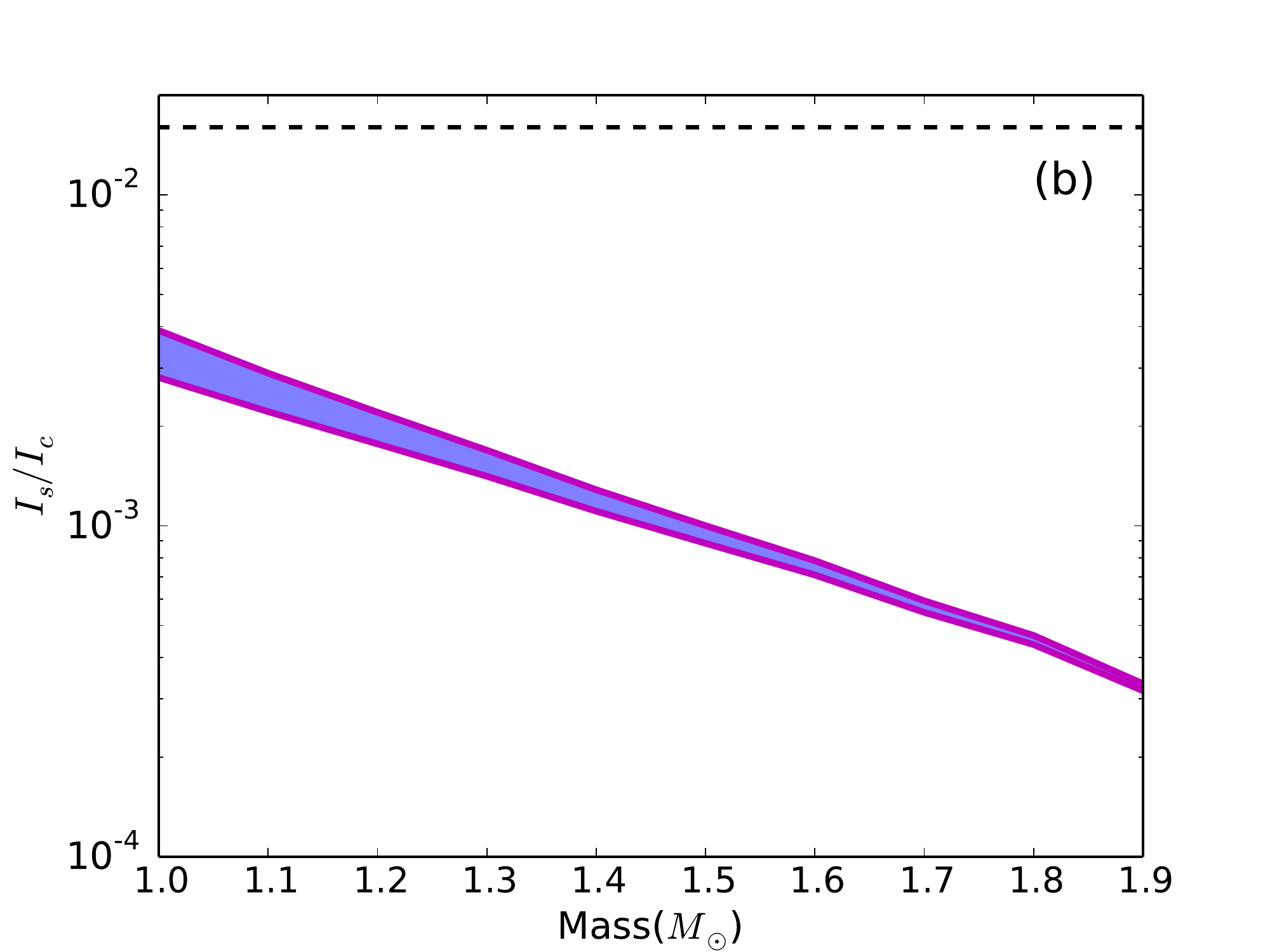}
\includegraphics[width=8cm,height=6cm]{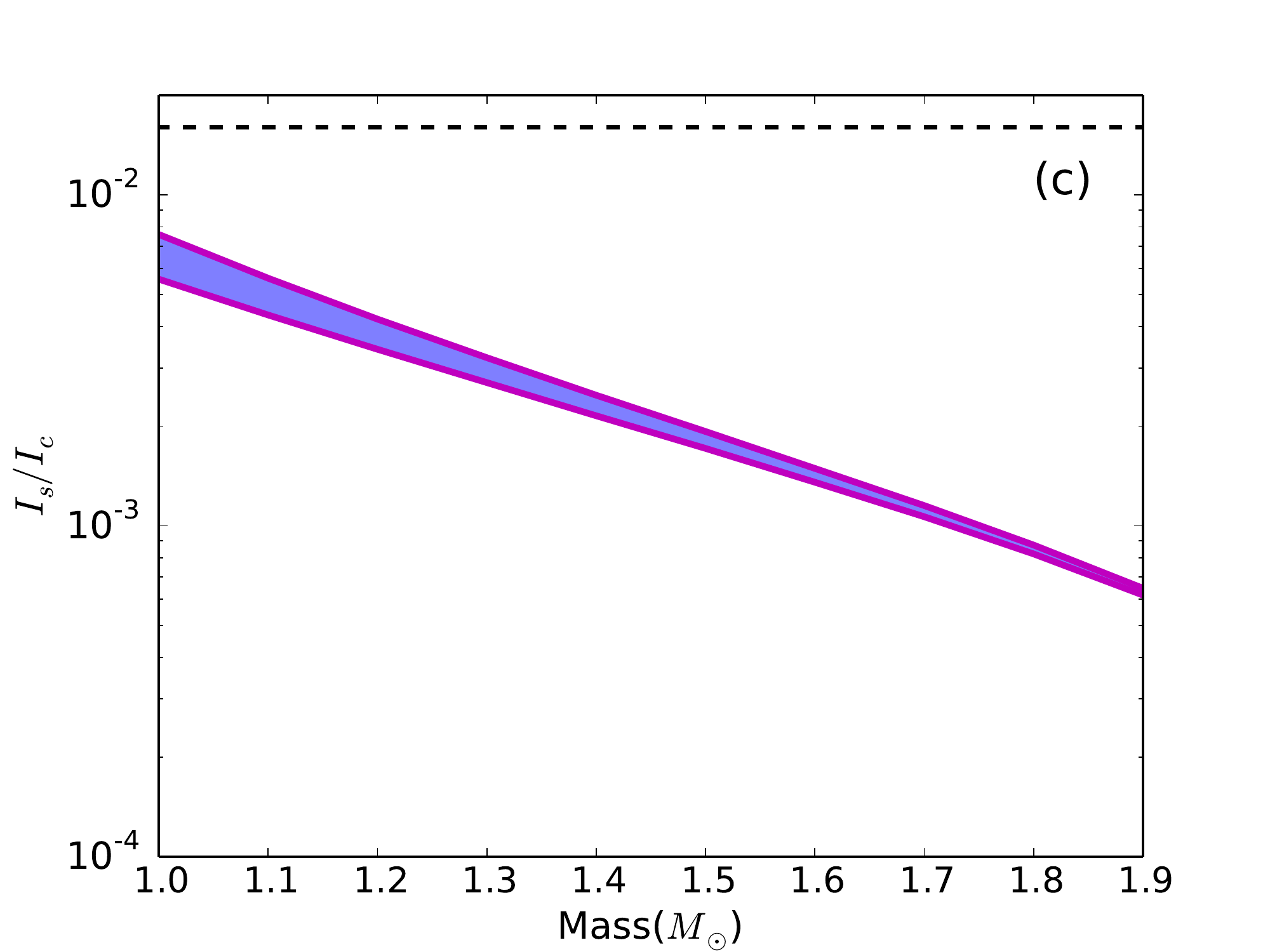}\hspace{0.5cm}\includegraphics[width=8cm,height=6cm]{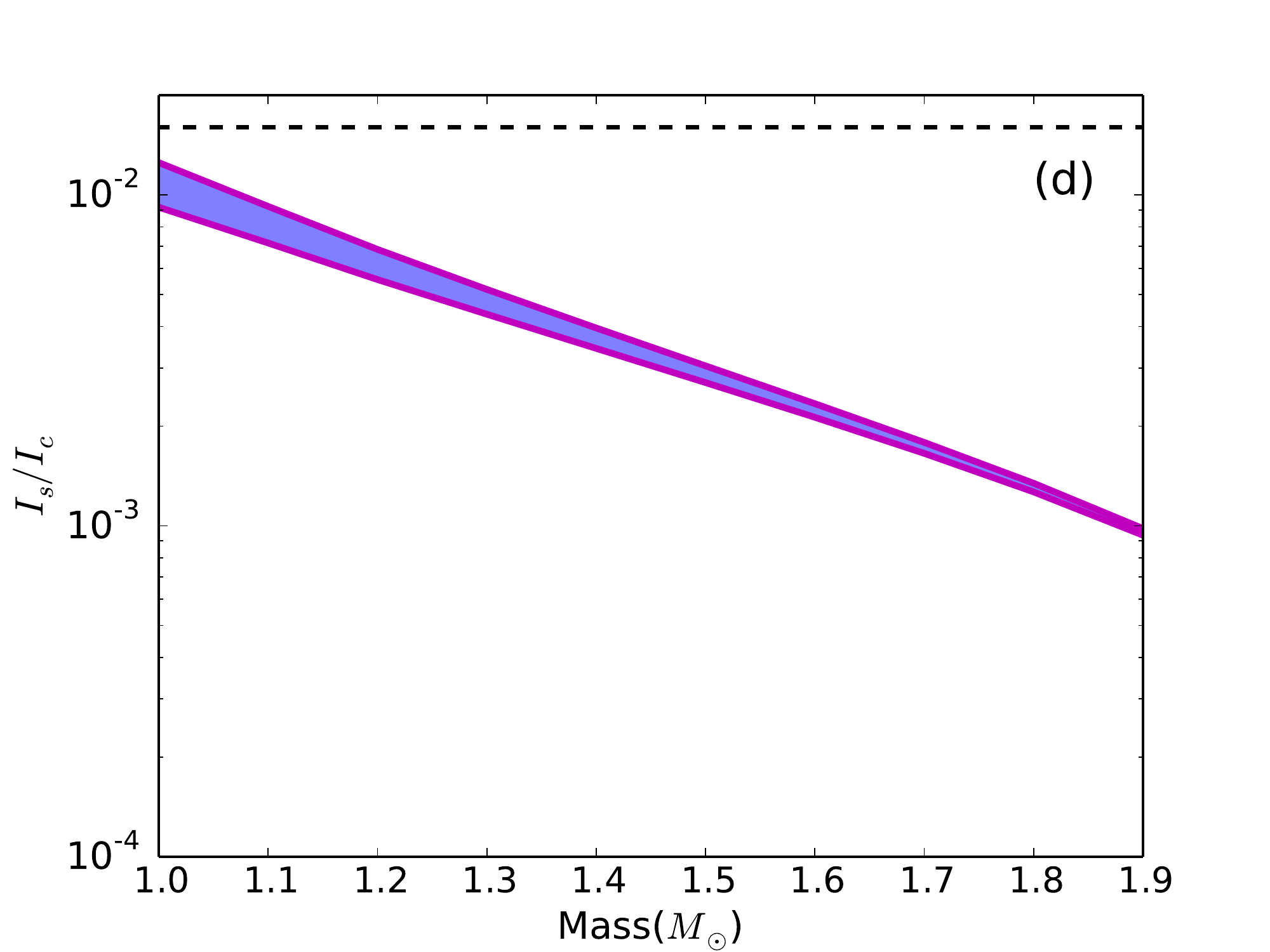}
\caption{Moment of inertia of the neutrons in the strong pinning region of the crust $I_{\rm s}$ relative to that of the component of the star coupled to the crust during the glitch $I_{\rm c}$ as a function of stellar mass for $L=30$MeV (a), $L=60$MeV (b), $L=90$MeV (c) and $L=120$MeV (d). The ranges correspond to taking the ranges of $I_{\rm c}$ shown in figure~\ref{fig:Ic/It}. The lower bound inferred from Vela's glitch activity is shown by the horizontal dashed line.}
\end{center}
\end{figure*}

As the slope of the symmetry energy $L$ increases (gets stiffer), the strength of the mutual friction $\bar{\mathcal{B}}(\tilde{r})$ gets larger throughout the core of the star. A larger $L$ leads to the symmetry energy as a function density of $E_{\rm sym}(\rho)$ that increases more rapidly, leading to a larger proton fraction at a given density, which contributes to an increasing $\bar{\mathcal{B}}(\tilde{r})$ with density (since $\mathcal{B}$ is almost proportional to $x_{\rm p}$). In addition, the microscopic effective mass $m_{\rm p}^*$ generically decreases with increasing density (from 0.7-0.85 at the crust down to 0.15-0.3 in the core for a 1.4$M_{\odot}$ star). Since $\mathcal{B}$ depends on $m_{\rm p} - m_{\rm p}^*$ and inversely on $m_{\rm p}^*$, this also contributes to an increasing $\bar{\mathcal{B}}(\tilde{r})$ with density. For the softest EOS shown, $L = 30$ MeV, the symmetry energy begins decreasing with density just above saturation density, and therefore so does the proton fraction. This behavior dominates the behavior of $\mathcal{B}$ at high density, which explains its non-monotonic behavior for $L = 30$ MeV.

At the softer end of the range of $L$ we consider, the mutual friction profile changes rapidly with variations in $L$. For the softest EOS shown, $L=$30 MeV, the non monotonic behaviour arises because the proton fraction decreases with increasing density from the crust to the core, while the effective proton mass increases. For stiffer EOSs, both quantities increase from crust to core.

For $L=30$ MeV the predicted mutual friction profile lies almost completely within the inferred range. The upper end of the range of $\bar{\mathcal{B}}(\tilde{r})$ inferred from observation, $1.3\times10^{-4}$, implies that the core neutrons are entirely decoupled from the crust at the time of glitch, and that the superfluid neutrons in the strong pinning region in the crust have only the rest of the crust and the core protons to spin up during the glitch event. At the lower end of the range, for a canonical 1.4 M$_{\odot}$ neutron star a small portion of the outer core becomes coupled to the crust. For higher masses an increasing part of the inner core also becomes decoupled.

If, on the other hand, we consider the lower end of the inferred range of $\bar{\mathcal{B}}(\tilde{r})$, $3\times10^{-5}$, most of the core (within the inner $\approx 10$km) is coupled to the core. For $L$=60-120 MeV, most of the core has mutual friction significantly stronger than the inferred range, and hence a large fraction of core neutrons are coupled to the core at the time of the glitch, regardless of whether we consider the upper or lower bounds of the range.

In Fig.~5 we plot the ratio of the moment of inertia of the coupled part of the core $I_{\rm c}$ (that which has a mutual friction parameter larger than the observationally inferred values) to the total moment of inertia of the star $I_{\rm t}$ as a function of mass $M$ for the $L$=30,60,90 and 120 MeV members of the SkIUFSU EOS. The shaded area in each plot spans the upper and lower bounds for $I_{\rm c}/I_{\rm t}$ corresponding the fractions $I_{\rm c}$ obtained from considering the upper and lower bounds of the inferred range of $\bar{\mathcal{B}}(\tilde{r})$, $3\times10^{-5}$ as outlined above in reference to Fig.~4. For $L$=30 MeV, the fact that the predicted mutual friction profile lies mostly within the inferred range of $\bar{\mathcal{B}}(\tilde{r})$ means that the ratio $I_{\rm c}$ is almost completely unconstrained by the post-glitch evolution - as can be seen, between almost 0\% and 100\% of the core could be coupled to the crust during the glitch rise time. For intermediate-to-stiff symmetry energy behaviors at saturation density, $L=60-120$ MeV, the range of $I_{\rm c}/I_{\rm t}$ is inferred to be above 70\% (above 80\% for star $>1.4M_{\odot}$).

%
%
\begin{figure*}\label{fig:IUFSU}
\begin{center}
\includegraphics[width=8cm,height=6cm]{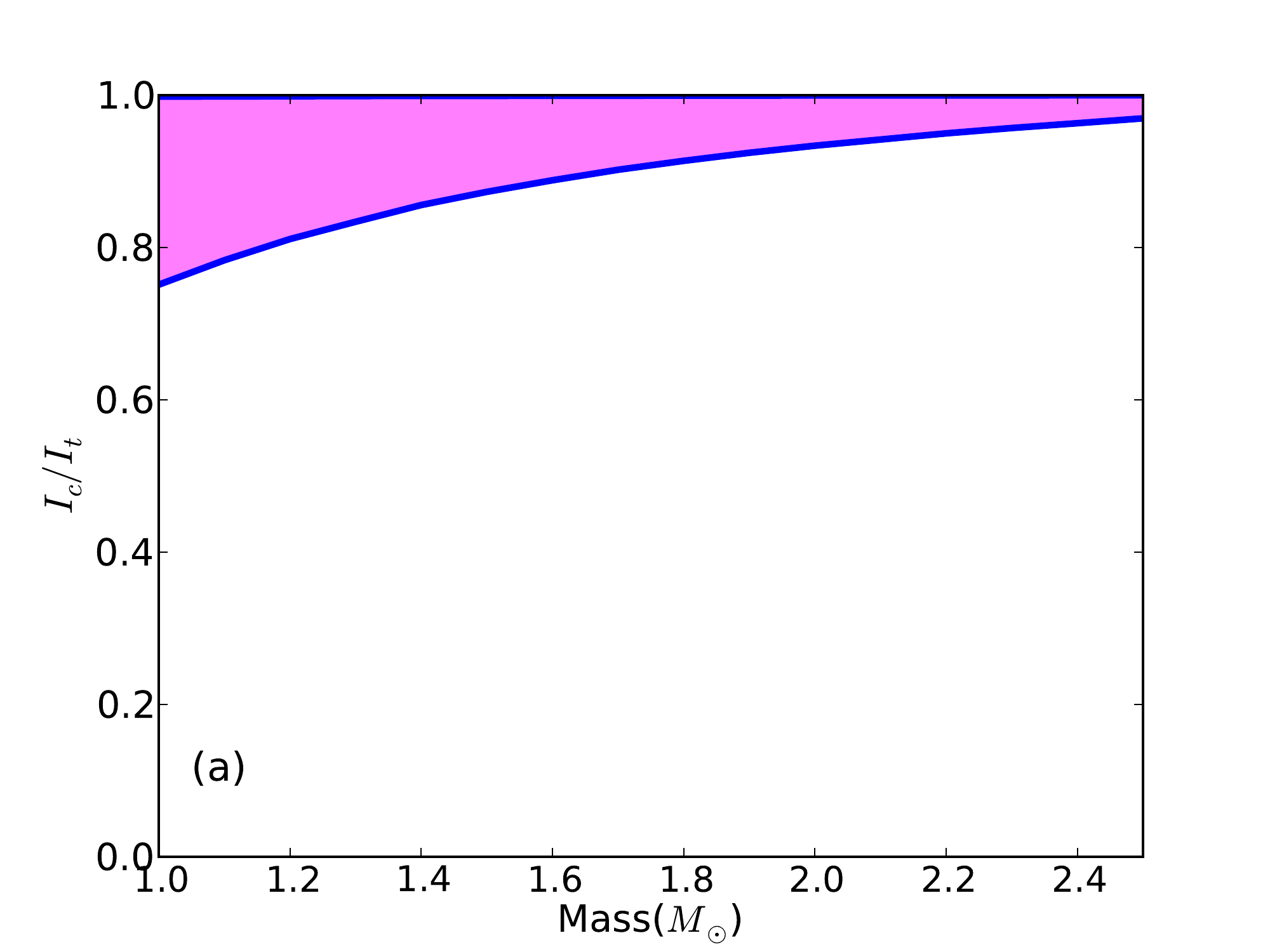}\hspace{0.5cm}\includegraphics[width=8cm,height=6cm]{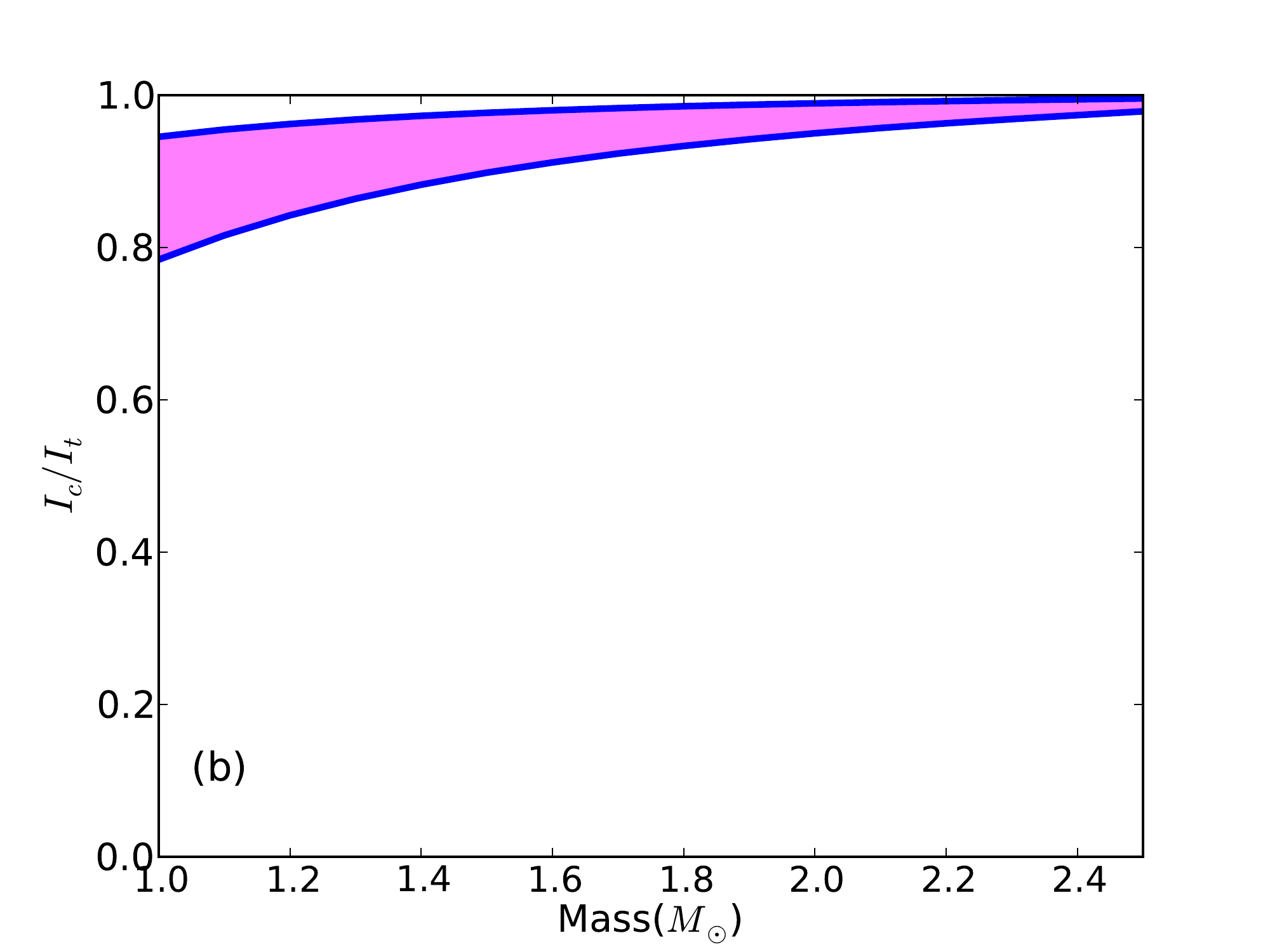}
\includegraphics[width=8cm,height=6cm]{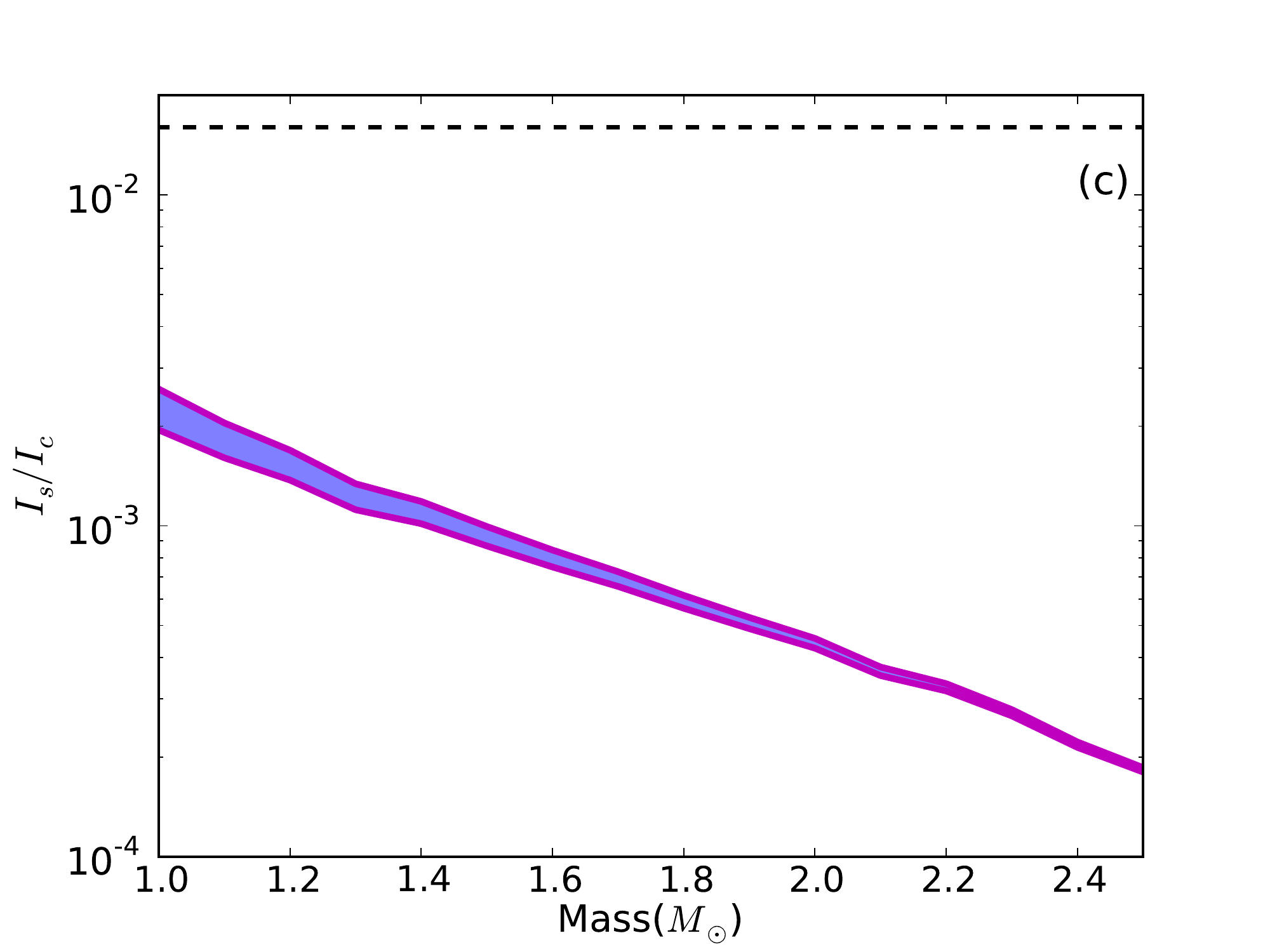}\hspace{0.5cm}\includegraphics[width=8cm,height=6cm]{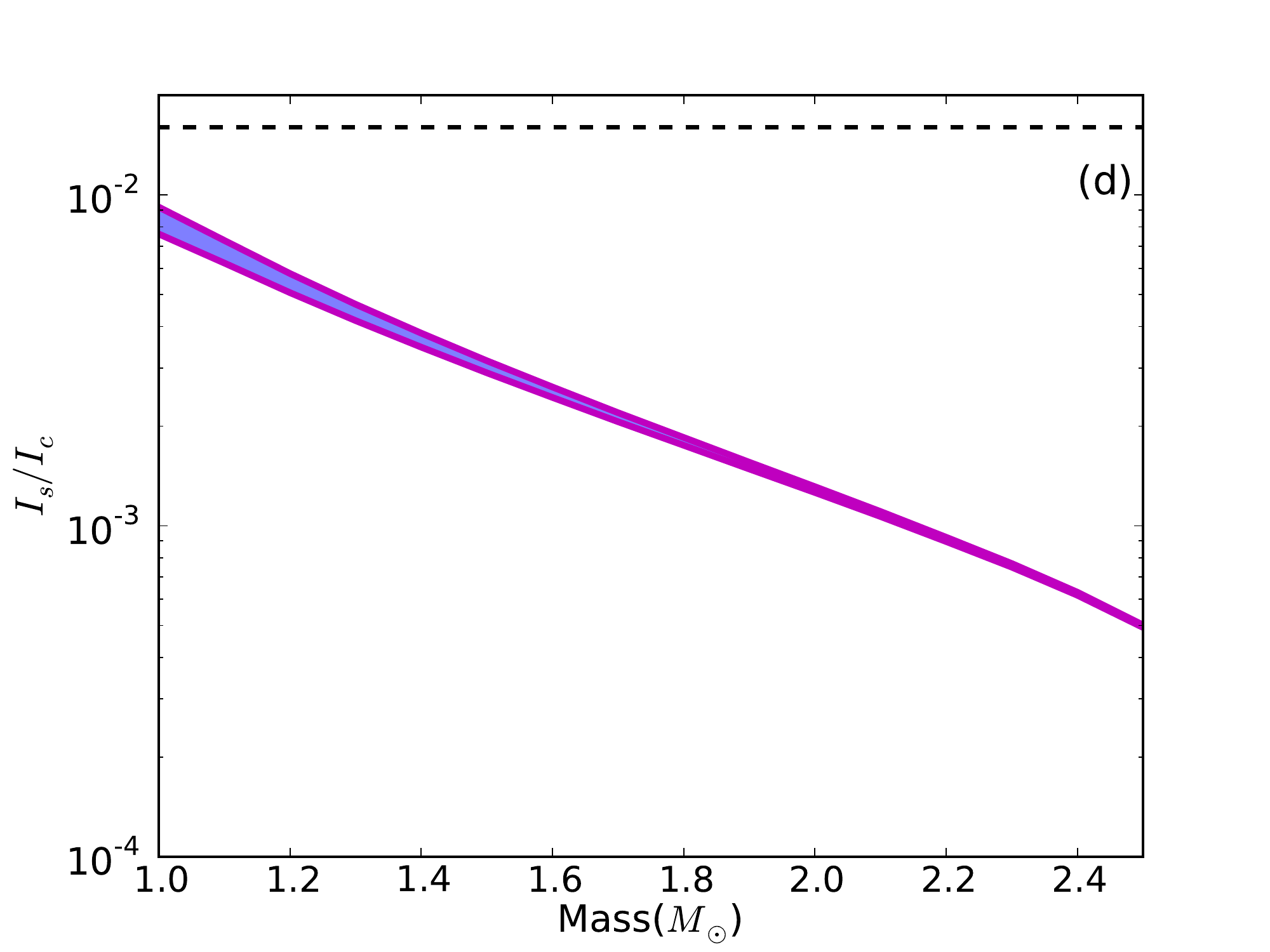}
\caption{Fractional moment of inertia of the component of the star coupled to the crust during the glitch $I_{\rm c}/I_{\rm tot}$ (a,b) and moment of inertia of the neutrons in the strong pinning region of the crust $I_{\rm s}$ relative to that of the component of the star coupled to the crust during the glitch $I_{\rm c}$ (c,d) for the $L=30$MeV (a,c), $L=90$MeV (b,d) members of the IUFSU family of EOSs. The bands have the same meaning as figures 5 and 6.}
\end{center}
\end{figure*}

%
%
\begin{table}\label{Tab:n+}
\caption{Extension of the strong pinning region into the outer core required to satisfy the Vela constraint on $G$ for a 1.4$M_{\odot}$ neutron star for $L=30$ MeV and different values of $Y_{\rm g}$. The extension is given as the baryon number density above the crust core transition density, $n_{\rm +}$.} \label{Table1}
\begin{tabular}{lclclcl}
\hline
$L$(MeV) & $n_{\rm +}$ ($\mathcal{B}$ = 1.3$\times10^{-4}$) & $n_{\rm +}$ ($\mathcal{B}$ = 3$\times10^{-5}$)  \\
\hline
\hline
30                 &   0.016                        &      0.083   \\
60                 &   0.048                        &      0.051   \\
90                 &   0.032                        &      0.035   \\
120                 &   0.023                        &      0.025 \\
\hline
\end{tabular}
\end{table}

Given the moment of inertia of the predicted fraction of the core of the neutron star $I_{\rm c}$ coupled to the crust obtained as detailed above, we can calculate the ratio of the moment of inertia of the neutrons in the strong pinning region, assumed for now to be entirely within the inner crust, to the moment of inertia of the couple core fraction $I_{\rm s}/I_{\rm c}$. In Fig.~6 we compare this value as a function of mass to the minimum required to explain the Vela glitch activity over the past 45 years $G=0.016$, indicated by the horizontal dashed line in all four plots. It is important to note here that this comparison is meaningful only if the observed sizes of the glitch are a reasonable approximation to the actual change in frequency after the rise. This may not be the case if, as some simulations suggest \citep{Haskell:2011xe}, the rise time is significantly shorter than the current upper limit of $\approx 1$ minute and a larger fraction of the core may be decoupled during a glitch. Furthermore the comparison is appropriate for the Vela pulsar, for which the activity is dominated by giant glitches where we can expect Mutual Friction to dominate the recovery. This is unlikely to be the true for other pulsars in which smaller glitches are frequent and are unlikely to lead to lag reversal.

Returning to Fig. ~6, the first thing to notice is that for no EOS in the SkIUFSU family within the wide range of $L$ considered does the predicted value of $I_{\rm s}/I_{\rm c}$ exceed the required value to account for Vela glitches. Within this family of EOSs, the crustal neutrons are insufficient to explain the Vela glitches, using a core fraction consistent with the initially observed glitch recovery.

Generically, the ratio $I_{\rm s}/I_{\rm c}$ decreases with increasing mass as the crust becomes thinner while the radius stays relatively constant (refer to Fig.~\ref{fig:MR}). At lower $L$ the predicted range for $I_{\rm s}/I_{\rm c}$ spans an order of magnitude owing to the wide range of possible coupled core fractions. The upper bound of the range corresponds to the lower bound of $I_{\rm c}/I_{\rm t}$, which in the case of $L=30 MeV$ is very small as the core is completely uncoupled over the glitch rise time. Because of this, the upper bound comes closest to matching the required level from the Vela pulsar glitches. In the three other cases, between 70 and 100\% of the core is coupled, and the range predicted for $I_{\rm s}/I_{\rm c}$ increases with $L$ as the star's crust increases; it does not rise enough to match the required level, though.

If one allows the region where the neutrons experience strong pinning to penetrate the core to an arbitrary amount, $I_{\rm s}$ can be increased until the ratio $I_{\rm s}/I_{\rm c} > 0.016$. In Table~2 we give the density \emph{above} the crust-core transition density to which we must extend the inner boundary of the strong pinning region in order to satisfy the Vela constraint for a $1.4M_{\odot}$ star. Pinning could occur in the outer core, for example, on magnetic fluxtubes. 

We next explore the extent to which changing the high density EOS changes the outcomes of our predictions. We now use the IUFSU family of EOSs with the parameter $\zeta=0.00$ which leads to the stiffest EOS at super-saturation densities and a corresponding maximum mass of $2.6M_{\odot}$. The EOS and symmetry energy is the same as the SkIUFSU up to saturation density. We show representative results in Fig.~7. In the left-hand panels we show the predicted ranges for the moment of inertia of the coupled core fractions to the total moment of inertia $I_{\rm c}/I_{\rm t}$, while in the right-hand panels we show the ratio of the moment of inertia of the neutrons in the strong pinning region to the moment of inertia of the couple core fraction $I_{\rm s}/I_{\rm c}$, compared to the minimum value inferred from the Vela pulsar. We show results for a representative soft EOS $L=30$MeV and representative stiff EOS $L=90$MeV from the IUFSU $\zeta=0.00$ family. The increased stiffness of the high-density EOS prevents the proton fraction dropping and weakening the mutual friction for $L=30$MeV as happens with the corresponding EOS from the SkIUFSU family. The behavior of the coupled core fraction is thus similar to that of the stiffer EOSs including the $L=90$MeV shown, above $\approx$80\% for all masses.

\section{Discussion and Conclusions}

Evidence exists for a short timescale ($\sim 1$ min) component in the post-glitch recovery of the 2000 and 2004 Vela glitches, suggestive of the recoupling of that portion of the stellar interior that was decoupled from the crust during the glitch. Under the assumption that the recoupling is mediated by the mutual friction force which results from electron scattering off the magnetized cores of superfluid vortices, we use the observed timescales and magnitudes to infer the moment of inertia of the region of the core coupled to the crust during the glitch.

Within the framework of hydrodynamic models of vortex dynamics, which suggest the pinning region may be confined to the region of the inner crust where the vortices are completely immersed in the crustal lattice, we calculate the ratio of the moment of inertia of the pinning region of the crust to the moment of inertia of  the core coupled to the crust during the glitch which we compare with the observed Vela glitch activity. We derive the neutron star EOS from the crust to the core, including the composition, microscopic effective proton mass and crust-core transition density, consistently using the same underlying nuclear matter EOS. We choose two families of EOSs: one derived from the SkIUFSU parameterization of the phenomenological non-relativistic Skyrme model of the in-medium neutron-proton interactions and one derived from the IUFSU parameterization of the phenomenological relativistic mean-field (RMF) model. The Skyrme family of EOSs give maximum mass neutron stars of $\approx 2.0M_{\odot}$, while the RMF family give maximum mass neutron stars of $\approx 2.6M_{\odot}$. Each family is generated by systematically varying the slope of the symmetry energy at saturation density over a conservative range $L=30-120$MeV, thus exploring the impact on our results of the current uncertainty in the EOS of neutron rich matter at saturation density.

The main conclusion of our analysis is that taking the strong pinning region to be confined entirely to the crust, no EOS results in large enough glitches to explain the Vela pulsar's glitch activity for neutron stars above $1 M_{\odot}$. However, for the stiffest EOSs considered, neutron stars with masses slightly under $1 M_{\odot}$ would marginally satisfy Vela's glitch activity, and uncertainties on glitch sizes and rise times could lead to lower fractions of the core being coupled during a glitch and make low masses consistent with the data.

We find that increasing the stiffness of the high density EOS, has only a small effect on the results for 1.4$M_{\odot}$ neutron stars, generally increasing the moment of inertia coupled to the crust during the glitch slightly. This high density stiffening has two effects: increasing the radius of the star, and thus the moment of inertia of that part of the core that must be spun-up, and also increasing the crust thickness, and therefore the angular momentum reservoir which spins the star up. These two effect oppose each other, resulting in only a small effect overall.

It would thus appear that even if one allows for part of the core superfluid to decouple during a glitch it is quite generally the case that, if standard mutual friction mechanisms are at work, strong entrainment in the crust limits the amount of angular momentum available for a glitch to below the level required to explain the glitch activity of the Vela pulsar. Allowing the strong pinning region to extend by up to $0.05$fm$^{-3}$ into the core, would allow all EOSs to produce glitches consistent with Vela's glitch activity. Physically this would correspond to the protons in the outer core being in a type-II superconducting state and the superfluid neutron vortices being pinned to superconducting flux tubes \citep{Link:2003hq, Haskell:2012qk}, and contributing to the angular momentum reservoir available for a glitch. Note, however, that in this case vortex-flux tubes interactions could be the leading source of Mutual Friction in some areas of the core \citep{Sidery:2009al, Haskell:2014rm}, thus invalidating our assumption that Mutual Friction is due to electron scattering off vortex cores. 

The strength of mutual friction throughout the core increases with increasingly large symmetry energy slopes $L$. At low values of $L$ (soft EOSs), the fraction of the core coupled to the crust at the time of the glitch is quite unconstrained by the observations. At intermediate-high values of $L$ (stiffer EOSs), the fraction of core coupled to the crust is constrained by observations to be above $\approx$70\% for realistic neutron star masses.

The values extracted for the coupled fractions are contingent not only on our model for the crust-core coupling mechanism, but also the robustness of the observational data. As such, our results are tentative, as they depend on only two observations, one of which (the 2004) is particularly marginal. The fitting of exponential recovery terms can also be called into question, as on short timescales the frequency evolution may be described by a very different functional form \citep{Haskell:2011xe, Link2014a}. It is important to note that our conclusions also rely heavily on reliability of estimates of the frequency after a glitch. This may not be the case if the glitch rise time is significantly shorter than the current upper limit of $\approx 1$ minute, e.g. closer to timescales of a few seconds as simulations suggest \citep{Haskell:2011xe}, and a significantly larger fraction of the core may be decoupled during a glitch. If we allow all the core neutrons to be entirely decoupled during the glitch (requiring coupling timescales of less than 5-10s depending on the EoS), we still do not obtain consistency with a 1.4$M_{\odot}$ Vela pulsar except for the stiffest EOS, $L=$120 MeV. However, given a reasonable uncertainty of order 0.01 fm$^{-3}$ in the crust-core transition density \citep{Newton:2011dw} at a given value of $L$, our model would be consistent with Vela being a $M=1.4_\odot$ neutron star if $L\gtrsim70$ MeV and the coupling timescale is sufficiently fast to decouple $\gtrsim70\%$ of the core neutrons. From our calculations, this requires the coupling timescale to be $\approx$ 10s or less, as opposed to the minute timescales estimated from the observed glitch recoveries. Consistency with Vela being a $M=1.0_\odot$ neutron star can be achieved for $L\gtrsim50$ MeV, provided the coupling timescale is $\approx 30s$ or less, enough to decouple $\gtrsim80\%$ of the core neutrons.

The one aspect of our analysis that is not consistent with the underlying EOS is the use of equation~\ref{eqn:B} for the mutual friction coefficient; although the proton fraction and proton effective mass are derived from the underlying EOS, the form of the equation is a fit to a single microscopic calculation. Although we expect the basic functional dependence on quantities not to change much, the detailed form may depend on the EOS. One should also note that the values for the mesoscopic effective neutron masses in the crust, which encode the entrainment effect, are also not consistent with the EOS used, and could substantially alter the estimates of the angular momentum available in the crustal superfluid. Another important assumption of our model is that the protons and electrons in the core are coupled to the crust on short timescales by the magnetic field. This assumption, however, depends strongly on the magnetic field geometry and may not hold for all regions of the core \citep{KGPL15}.

Nevertheless, it is likely that the post-glitch response of glitches bears the imprint of the crust-core coupling dynamics, and our study illustrates one way in which such information can be extracted.

\section*{ACKNOWLEDGEMENTS}
The authors thank Pierre Pizzochero for helpful comments on the manuscript.
This work is supported in part by the National Aeronautics and Space Administration under grant NNX11AC41G issued through the Science Mission Directorate and the National Science Foundation under grants PHY-0757839 and PHY-1068022 and REU grant 1359409. BH is supported by the Australian Research Council via a DECRA fellowship.

\bibliographystyle{mn2e}

\end{document}